\documentclass[12pt,preprint]{aastex}

\shorttitle{The Outer Limits Survey}
\shortauthors{Saha et al.}

\begin{document}


\title{First Results from the NOAO Survey of the Outer Limits of the Magellanic Clouds}


\author{Abhijit Saha\altaffilmark{1,2}}
\affil{National Optical Astronomy Observatory, Tucson, AZ 85719}
\email{saha@noao.edu}

\author{Edward W. Olszewski\altaffilmark{2}}
\affil{Steward Observatory, The University of Arizona, Tucson, AZ}
\email{eolszewski@as.arizona.edu}

\author{Brian Brondel\altaffilmark{2,3}}
\affil{Department of Astronomy, Indiana University, Swain West 319, 727 East Third Street, Bloomington, IN 47405-7105}
\email{bbrondel@gmail.com}

\author{Knut Olsen\altaffilmark{1,2}, Patricia Knezek\altaffilmark{1,2}, Jason Harris\altaffilmark{1,2}}
\affil{National Optical Astronomy Observatories, Tucson, AZ 85719}
\email{kolsen@noao.edu, pknezek@noao.edu, jharris@30doradus.org}

\author{Chris Smith}
\affil{Cerro Tololo Inter-American Observatory, National Optical Astronomy Observatory, Casilla 603, La Serena, Chile}
\email{csmith@ctio.noao.edu}

\author{Annapurni Subramaniam\altaffilmark{2}}
\affil{Indian Institute of Astrophysics, Koramangala II Block, Bangalore-34, India}
\email{purni@iiap.res.in}

\author{Jennifer Claver}
\affil{National Optical Astronomy Observatory, Tucson, AZ 85719}
\email{jclaver@noao.edu}

\author{Armin Rest\altaffilmark{4,6}}
\affil{Cerro Tololo Inter-American Observatory, National Optical Astronomy Observatory, Casilla 603, La Serena, Chile}
\email{rest@ctio.noao.edu}

\author{Patrick Seitzer}
\affil{Department of Astronomy, University of Michigan, Ann Arbor, MI 48109}
\email{pseitzer@umich.edu}

\author{Kem H. Cook}
\affil{IGPP, Lawrence Livermore National Laboratory, MS L-413, P.O. Box 808, Livermore, CA 94550}
\email{kcook@llnl.gov}

\author{Dante Minniti\altaffilmark{5}}
\affil{Department of Astronomy, Pontificia Universidad Catolica de Chile, Av. Vicu–a Mackenna 4860, 782-0436 Macul, Santiago, Chile}
\email{dante@astro.puc.cl}

\and

\author{Nicholas B. Suntzeff}
\affil{Department of Physics, Texas A\&M University, 4242 TAMU, College Station, TX77843}
\email{suntzeff@physics.tamu.edu}

\altaffiltext{1}{NOAO is operated by the Association of Universities for 
Research in Astronomy, Inc. (AURA) under cooperative agreement with the 
National Science Foundation}
\altaffiltext{2}{Visiting astronomer, Cerro Tololo Inter-American Observatory, 
National Optical Astronomy Observatory, operated by AURA, Inc.\ under contract to the National Science
Foundation.}
\altaffiltext{3}{Current address: Astronomical Consultants \& Equipment Inc, PO Box 91946, Tucson, AZ 85752-1946}
\altaffiltext{4}{Physics Department, Harvard University, 17 Oxford Street, Cambridge, MA 02138}
\altaffiltext{5}{also at Vatican Observatory, V00120 Vatican City State, Italy}
\altaffiltext{6}{Space Telescope Science Institute, 3700 San Martin Drive, Baltimore, MD 21218}


\begin{abstract}

We describe the first results from the Outer Limits Survey, an NOAO
survey designed to detect, map, and characterize the extended
structure of the Large and Small Magellanic Clouds.  The survey
consists of deep images of 55 $0.6^{o} \times 0.6^{o}$ fields
distributed at distances up to 20$^{o}$ from the Clouds, with 10
fields at larger distances representing controls for contamination by
Galactic foreground stars and background galaxies.  The field
locations probe the outer structure of both the LMC and SMC, as well
as explore areas defined by the Magellanic Stream, the Leading Arm,
and the LMC orbit as recently measured from its proper motion.  The
images were taken with C, M, R, I, and DDO51 filters on the CTIO
Blanco 4-m telescope and Mosaic2 camera, with supporting calibration
observations taken at the CTIO 0.9-m telescope.  The CRI images reach
depths below the oldest main sequence turnoffs at the distance of the
Clouds, thus yielding numerous probes of structure combined with good
ability to measure stellar ages and metallicities.  The M and DDO51
images allow for discrimination of LMC and SMC giant stars from
foreground dwarfs, allowing us to use giants as additional probes of
Cloud structure and populations.

From photometry of 8 fields located at radii of 7$^{o}$ to 19$^{o}$
north of the LMC bar, we find main sequence stars associated with the
LMC out to 16$^{o}$ from the LMC center, while the much rarer giants
can only be convincingly detected out to 11$^{o}$.  In one field,
designated as a control, we see the unmistakable signature of the
Milky Way globular cluster NGC 1851, which lies several tidal radii
away from the field center.  The color-magnitude diagrams show that
while at 7$^{o}$ radius LMC populations as young as 500 Myr are present,
at radii $\gtrsim11^{o}$ only the LMC's underlying old 
metal-poor ([M/H]$\sim-1$) population remains, demonstrating the
existence of a mean population gradient at these radii. Nevertheless, even at 
extreme large distances, the dominant age is significantly younger than that 
of the Galactic globular clusters.  The
main-sequence star counts follow an exponential decline with distance
with a scale length of 1.15 kpc, essentially the same scale length as gleaned for
the inner LMC disk from prior studies.  While we cannot rule out the
existence of undetected tidal features elsewhere in the LMC periphery,
the detection of an ordered structure to 12 disk scale lengths is
unprecedented, and adds to the puzzle of the LMC's interaction history
with the SMC and the Milky Way.  Our results do not rule out the
possible existence of an LMC stellar halo, which we show may only
begin to dominate over the disk at still larger radii than where we
have detected LMC populations.
 
\end{abstract}





\section{Introduction, Motivation and History}
\label{sec_intro}


In our Galaxy, the most metal poor and (plausibly) the oldest stars
are distributed in a halo that extends beyond 25 kpc. Their spatial
distribution, chemical composition and kinematics provide clues about
the Milky Way's early history, as well as its continued interaction
with neighboring galaxies.  
In structure evolution models, the Clouds ought to have similar but scaled 
down accretion histories as $L^{*}$ galaxies, and ought to have dark  (and presumably also stellar)
halos.  Interaction with the Milky Way should begin to strip away these halos, with a 
rate depending on the ratio of masses of the galaxies, on the perigalacticon distances, 
and the time spent in the ``tidal region''.  Thus questions like the following must tell us 
about the formation and interaction histories of the Clouds.
 How old are the oldest stars in the extremities
of the Clouds?  How far do such stellar distributions extend? 
What tidal structure is revealed? Is there a continuity in the stellar 
distribution between the LMC and SMC?  Do they share a common 
halo with the Galaxy?  What do the kinematics of stars in outlying 
regions tell us about the dark matter distribution? Do they shed any light 
on the orbital histories of the Clouds? 

2MASS and DENIS are examples of panoramic surveys which have
yielded fundamental information about the LMC disk: \citet{vdm01},
using star counts of Red Giant Branch (RGB) and Asymptotic Giant
Branch (AGB) stars, derived the tilt of the LMC disk against the sky,
its intrinsic elongation, and its scale length.  These probe stars
fall off in density with distance from the LMC center, and eventually
become too sparse to be usable: thus the structure beyond 10$^\circ$, 
which is well within the LMC's estimated tidal radius  \citep{wein00}, 
remains unknown.  As a result, much of the
extant evidence for an LMC halo is indirect.  \citet{fea68} found that
velocity dispersions of planetary nebulae exceeded that of {\rm HII} regions,
suggesting a spheroidal component. \citet{schom92} demonstrated the
same from the kinematics of old clusters, but also argued that the
dispersion ($\sim 30\ {\rm km~s}^{-1}$) is not large enough for an
isothermal halo, and that the old cluster kinematics are consistent with disk rotation.
The inner RR Lyrae stars define a disk with characteristics similar to those 
obtained from AGB stars and Cepheids \citep{sub09}.
 \citet{min03} obtained velocity
dispersions of RR Lyrae stars distributed {\em over the bar} of $53 \pm 10\ {\rm
km~s}^{-1}$, which has been taken as evidence that a halo exists. However, 
given that these reflect conditions over the projected center of the galaxy,
it is not clear whether we are seeing a full fledged halo, or just the central bulge. 

\citet{wein00} showed that, assuming the older LMC space
velocity, that LMC disc orbits are affected by the tides caused by
the Milky Way. These tides thickened the disk and created a set of `halo'
objects rotating with the plane, thus explaining the results of \citet{schom92} 
and \citet{fre83}.

Direct detection of spatially extended structure in the LMC/SMC is
difficult, because of the considerable extent on the sky that such an
entity would occupy, and because the LMC bar and disk are relatively face
on.  \citet{irw91} counted stars over an extended area from UKSTU
plates: his isopleths show that there is a steep decline in stars near
8$^\circ$ from the LMC and near 4$^\circ$ from the SMC (suggesting a
termination of their respective disks) but that tenuous structure
persists over an apparent size of $23 \times 17$ degrees with `halos'
of LMC and SMC overlapping. \citet{kin91} counted extra-tidal RR
Lyraes around NGC~1841 and Reticulum (11.5$^\circ$ and 15$^\circ$ from
the LMC bar) and found them consistent with a King model with 22
RR~Lyrae stars per square degree over the LMC bar.  On the other hand,
while these RR~Lyrae have velocities and distances consistent with LMC
membership, \citet{sun92} and \citet{alv04} have argued that the
data are also consistent with an extended disk model.  \citet{alv04}
further noted that the best fit {\it exponential} fit to the radial
run of RR~Lyrae stars has a scale length that matches that ``of the
LMC's blue light'', i.e. of its {\it bona-fide} disk.
 
\citet{str84} found intermediate age blue stars in the color-magnitude 
diagram (CMD) as far out as 9.5$^\circ$ from the LMC center, near the
globular cluster NGC~2257.  A deep CMD 8$^\circ$ north of the LMC
center by \citet{gal04} also shows bright blue main sequence stars (in
addition to stars with fainter turn-off brightness  showing ages from 
8 to 13 Gyr), strongly suggesting the
continued presence of  young stars ( $\sim$~2-3 Gyrs), and thus of the disk.
\citet{dol01} show that a field in the outskirts of the SMC also has
a wide range of ages, from 2~Gyrs old to at least 9-12 Gyrs old.
These results taken together show that we have not even begun to
determine how far any Cloud related structure extends and how its
stellar content changes; and whether such structure is flat and
disk-like, or whether dominated by a spheroidal halo.

The survey presented here, which goes out to  fields as far as 20 degrees from the LMC and to 
$\sim$12 degrees from the SMC,
is extreme compared to the spatial positions of other deep
CMDs such as \citet{gal04}.  However, there are some other shallower studies to note:
\citet{nid07} and \citet{maj99, maj09}
claim detection of LMC red giant stars
out to at least 23$^\circ$ from the LMC center, with LMC-like velocities.
\citet{mun06}
see stars in the field of the Carina dwarf spheroidal
that they interpret to be from the LMC, at an angular distance of 22$^\circ$
from the LMC.

\citet{vdm02} analyze the carbon star velocities in the LMC samples of \citet{kun97b} and 
unpublished work by Hardy, Schommer, \& Suntzeff, taking care to account for the full effects of the LMC's space motion on the radial velocity distribution of the carbon stars.  They find that the LMC disk has a rotation curve amplitude of $50 \pm 16~ {\rm kms}^{-1} $, that more than half of the $ \sim 9 \times 10^9 M_{\sun}$  enclosed within $ \sim 9$  kpc is due to a dark halo, that the disk is thick and is tumbling at a rate $ {\rm d}i / {\rm d}t  = -103^{\circ} ~{\rm Gyr}^{-1}$.  They also measure a dynamical center and position angle of the kinematic line of nodes consistent with those expected from their geometric studies of the shape of the LMC disk \citep{vdmC01, vdm01}, lending support to their analysis.  \citet{ols07} compare the LMC's carbon star kinematics with those of its red supergiants 
\citep{mas03}  and HI \citep{kim03}, while employing the updated LMC proper motion measurement of \citet{kal06}.  They find a larger rotation curve amplitude of $ \sim 60~ {\rm kms}^{-1}$ for the carbon stars, with larger values still for the HI and red supergiants.  They also conclude that  $\sim 10 \%$  of the carbon stars are being stripped along with HI in tidal streamers.   In contrast,  the Magellanic streams, which are the most extreme examples of disturbances and flows, have shown no evidence to date of corresponding stars.  It is thus important to see if evidence for stripping can be found farther out from the LMC center, where, if anything, tidal effects are expected to be more pronounced.  

It is worth noting here that while the term `halo' has been used to label a tenuous
extended spheroidal component (as gleaned from diffuse surface brightness) around galaxies other than our own, it is only in our own Galaxy, that the dynamical behavior is also known.
There continues to be confusion about whether so-called halos around galaxies are just a continuation of their bulges (in which case the late type spirals should have progressively weaker halos), or whether 
they are determined by some other process.    Understanding the dynamical 
properties of halos can provide clues, but spectroscopy of faint stars in all but the nearest galaxies 
is beyond current reach.  This was another strong motivation in the design of this survey: finding 
objects that can be used as dynamical tracers in the outermost regions of the Clouds.

Given this current state of knowledge, we wish to address the following 
questions:

\begin{enumerate}

\item How far do the disk structures of the LMC and SMC extend?  

\item Are there stars in an extended spheroidal distribution (halo), as in 
the Milky Way?  $\Lambda$CDM cosmology predicts a spheroidal dark
matter halo for galaxies like the LMC. Is there an accompanying
stellar halo?  What, if any, relation binds a stellar halo to 
the dark matter halos posited by the $\Lambda$CDM models?
The LMC/SMC/MW system appear to be in the early stages
of a merger, or a near merger.  A detailed study of the
components of the individual galaxies will help us better interpret
the imprints of past mergers in the current universe.

\item How do ages and metallicities change with position within each Cloud? 
What is the distribution of the {\it oldest} stars along the line
between the LMC and SMC?  What does this distribution reveal about the
history of interaction of the two galaxies?  Ages and metallicities
from this survey will tag intermixing of LMC-SMC-Galaxy-halo
populations in the boundary regions.

\item Are there stars associated with the Magellanic Stream?  
Our survey re-addresses this issue with much higher sensitivity.

\item What do the dark matter halos of the Clouds look like?  The survey 
is designed to identify  {\it individual} red giants  associated with the Clouds from
foreground dwarfs  using gravity sensitive
photometry comparing $DDO51$ and $M$ (see \cite{gei84}, \citet{mor01} and references
therein), a method that works even better when the giants are more
metal poor than the foreground dwarfs in the Milky Way.  The giants
can serve later as kinematic probes for tracing the potential and how
it changes going between the Clouds and the Galaxy. This provides a probe
for the dark matter halo(s), whose presence is predicted by the
$\Lambda$CDM cosmology.

\end{enumerate}

\section{Survey Goals and  Observational Design}
\label{sec_design}

We have carried out deep imaging in selected fields within an extended
region around the LMC/SMC complex, up to radial distances of 18~kpc
in projection
from the LMC,  and $\sim$12 kpc from the SMC in order to examine the
disk and/or halo structure of the Clouds at larger radii than have
been reliably examined before.  We have also included pointings in
various spots within and outside both the `leading' and `trailing'
sections of the Magellanic Stream, to look for their elusive
\citep{Maj03} stellar content, which should be present if these
streams have tidal origin.


Main sequence stars are the most unambiguous tracers of any stellar
population, and photometric information in appropriate pass-bands can
reveal metallicities and ages.  The LMC/SMC are at just the right
distance to see substantial portions of the main sequence (MS) with
ground based wide field imaging (seeing limited to $ \sim 1$ arc-sec)
before contamination from unresolved background galaxies becomes
important.  Also, unevolved stars in this unambiguous portion of the
MS out-number the corresponding more luminous evolved giants by a
factor of $\sim 100$.  For these reasons, we chose to set up a 5
passband system that concentrates on the main sequence, while still
studying the giants.

The data from our survey, whose observational design is laid out in
the remainder of this section, are crafted to enable us to learn about
the early history of the LMC and SMC, and about their interactions
with each other and with the Milky Way, by focussing on  the 
questions listed at the end of \S~\ref{sec_intro}.

\subsection{Passband Selection}

The filter choice for this survey is a hybrid of   Landolt  $R, I$, and Washington $C, M$
\citep{can76} ,  with $DDO51$  \citep{gei84}.   
Their respective functions will be described in more detail in \S~\ref{sec_cmds}, 
 where we will show that color-magnitude diagrams from
this set allows us to i) discriminate metallicities to within a factor
of 2, ii) constrain ages to 20\%, even for ancient populations, iii)
map spatial densities, metallcities and ages over a much wider outer
expanse of the LMC/SMC complex than ever before, with much greater
sensitivity than previous studies, and iv) pick out red giants
associated at the distance of the Clouds by discriminating against the
plethora of foreground Galactic red dwarfs, so that future
spectroscopic observations can furnish kinematic and detailed chemical
composition data.

For most fields foreground reddening is modest ($E(B-V) \le 0.1$), and
can be obtained from dust maps \citep{sch98}
or even more direct means.  For the majority of our fields, which are quite far out from 
either the LMC or the SMC, extinction from within these galaxies is expected to be insignificant.
In any case, as we shall see later, the actual data and resulting color-magnitude diagrams in 
multiple bands effectively delimit the possible range of total extinction in a given field, 
if self-consistent comparison with isochrones is demanded.  In particular, note that since to zeroth 
order both $C$ and $M$ passbands track metallicity, such sanity checks  from $CMRI$
photometry are effective, even though metallicity is an unknown.

\subsection{Survey Requirements and Exposure Details}

Exposures times were calculated according to the following stipulations:

\begin{enumerate}
\item $C, R, \& I$ images  must reach reach at least 1.5 mags past the 
turnoff brightness of stars for an old globular cluster placed at the
distance of the SMC (the farther of the two Clouds), with S/N $\approx
20$ or better.  This corresponds to $R \approx 24.0$, $I \approx
23.5$, and $C \approx 24.5$.  This is required to enable identification of 
a section of the main sequence that is populated by stars of all ages.

\item The photometric S/N on the giant branch of a globular cluster at 
the distance of SMC must be $\approx 50$ or better. This drives
considerable attention to the containment of systematic errors in the
photometry, and is necessary if we are to be able to compare systematically 
between different fields, differentiate reddening effects and discriminate metallicities.

\item Photometry in $M$ and $DDO51$ should have S/N $\approx 50$ on 
the SMC giant branch, with no requirements for fainter magnitudes
(i.e. not planned for use on the main sequence stars at Cloud
distances).

\item Short exposures that do not saturate  AGB stars at the LMC 
distance must also be taken in all 5 passbands.

\end{enumerate}

To meet the above stipulations,   a large range of brightness must be covered.
To do so, the exposures in the various pass bands were made up as follows:
\begin{itemize}

\item   $R$ :   $3 \times 580s  + 1 \times 50s + 1 \times 10s ~~=~~ 1800s$  total

\item   $I$ :  $4 \times 585s  + 1 \times 50s + 1 \times 10s ~~=~~ 2400s$ total

\item   $C$ : $3 \times 1080s + 1 \times 300s + 1 \times 60s ~~=~~ 3600s$ total

\item   $M$ : $2 \times 120s + 2 \times 30s ~~=~~ 300s$ total

\item   $DDO51$ :  $2 \times 750s  + 2 \times 150s  ~~=~~ 1800s$ total

\end{itemize}
 Thus each field takes 2.75 hours of exposure time, and with pointing
 and readout overheads, a total time of about 4 hours.  The data
 were acquired in several observing runs using MOSAIC2 on the Blanco
 4m telescope at CTIO, spanning more than a two year period from
 August 2006, through December 2008, in addition to some data
 from a pilot program (with the same instrument and telescope)
 from October 2005.  Approximately 20\% of the total time was lost 
 to weather or inadequate seeing conditions.  Exposures were {\it not}
 dithered.  On occasion, when the same field was observed on
 different observing runs, there were small unplanned offsets in the
 pointing.

\subsection{Photometric Calibration}

Realizing that there are times when useful imaging can be obtained
even though pristine photometric conditions do not prevail, and also
because observing standard stars with MOSAIC2 is relatively inefficient, 
we chose to acquire auxiliary data to calibrate our
observations.  The 0.9m with the CFCCD imager on CTIO was used to
observe standard fields, as well as parts of each MOSAIC2 target area.
These observations were not on the same nights as the MOSAIC2 data.
Only 0.9m data obtained in photometric conditions (as gleaned {\it a
posteriori} from the photometry residuals of standard stars throughout
the night) are used.  Of the 22 nights allocated for the 0.9m telescope 
in the period between August 2006 
and December 2009, 14 were deemed photometric, with standard star rms residuals 
in all bands of at most 0.025 mag. These observations establish local sequences on
each MOSAIC2 field, and eliminate the need to observe each field with
MOSAIC2 in perfect photometric conditions.  On each night of observing
with the CFCCD, target object fields were interspersed with standard
star fields from \citet{lan92} \& \citet{lan83} that also contain stars that have been 
calibrated for Washington $C$ and $M$ (as detailed in \S~\ref{sec_calCFCCD}). 

\begin{figure}
\epsscale{.80}
\plotone{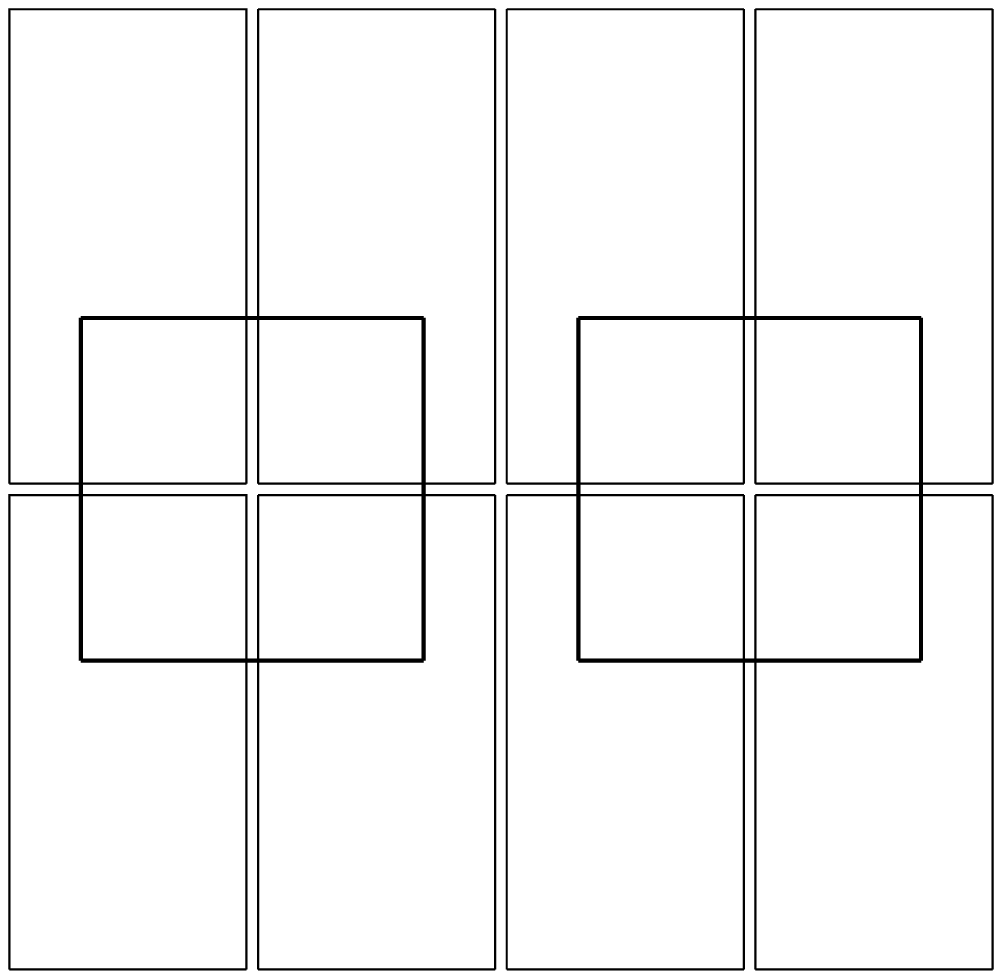}
\caption{A schematic diagram drawn to scale on the sky, showing
the relative placement of the eight 2K$\times$4K CCDs of  MOSAIC2
shown as faint rectangles, and of the two pointings on the CFCCD shown as 
the bold squares. With 2 pointings of CFCCD,  there are common stars between
CFCCD and each of the 8 CCDs of MOSAIC2. \label{fig_placement}}
\end{figure}

The CFCCD on the 0.9m covers a field area of 13 arc-minutes on a
side. Figure~\ref{fig_placement} shows the two positional placements
of the CFCCD with respect to the 8 chip format and area coverage of
the MOSAIC2 field-of-view (FOV). Thus for each field observed with
MOSAIC2, the plan calls for two placements of the CFCCD, which allows
secondary photometric sequences to be established on all 8 MOSAIC2
CCDs.  In practice there are photometric data in both
pointings for only half of the fields of
Table~\ref{tab_targfields}. However all  of the fields have at
least one CFCCD placement to establish the photometry.  Details of the
process, and measures of accuracy are deferred to
\S~\ref{sec_photometry}.

\subsection{Field Selection}

A complete survey around the LMC/SMC complex, covering from 8$^\circ$
to 20$^\circ$ distance from each of the galaxies, the `bridge' region
in between, and in and around the Magellanic Stream requires a
coverage of over 2000 square degrees.  Currently the most efficient
instrument complement available that can be pointed at the desired
region of sky is the combination of the CTIO-4m Blanco telescope and
the MOSAIC2 imager. A complete  survey is unfeasible, since
each pointing with MOSAIC2 (8 CCDs, each with $2K \times 4K$ pixels)
covers only  $36 \times 36$ square arc-min or $~0.36$ square
degrees of sky.  A `reasonable' survey program of 30 nights covers
50-60 pointings, at
least 10\% of which need to be on control fields.  This allows only
about a 1\% fill-factor of the region of interest, so the fields must
be chosen purposefully.  With the LMC and SMC interacting with each
other, and both interacting also with the Galaxy, we should not expect
much spatial symmetry. We must, at a minimum, observe along several
directions, each of which we expect to either be dominated by, or
least affected by key aspects of the LMC/SMC/Galaxy interactions, so
that we may attempt to disentangle them.

Accordingly we focussed attention on 5 regions: 
\begin{enumerate}

\item Looking away from the LMC in a direction least 
complicated by the SMC. This is along North from the LMC from about
$7^\circ$ to 20$^\circ$ from the LMC center, which is past the nominal
tidal radius of the LMC.  These fields are labelled F7N, F9N, F11N, F12p5N, F14N, 
F121, F122 and F123, with two flanking fields F111 and F113. 
Additional fields (designated F141, F142, F143 and F144) at distances from
$11^\circ$ to $17^\circ$  
degrees running NW from the LMC were also
chosen to sample more outlying fields with the original goal of
searching for tidally stripped stars predicted by \citet{wein00}.

\item Almost due North from the SMC, along the Magellanic stream. These
field designations begin with `F3'.

\item Towards the Galactic plane from the LMC along the `leading arm' 
of the Magellanic Stream \citep{put98}.  These field labels begin with `F4'.

\item  Away from the LMC/SMC complex, westwards from the SMC, orthogonal to the 
Stream, which is also an area opposite from the LMC and so least
affected by it. The names begin with `F5'.

\item  Several control fields 30$^\circ$ to 40$^\circ$ away from the Clouds, 
 spanning a range of Galactic latitudes that bracket the levels of
 foreground contamination in our target fields. This is necessary to
 make good models to account for contamination.  These field names begin 
 with 'C'.

\item In a direction looking back along the LMC's  trajectory, based on 
the proper motion studies by \citet{kal06}  and \citet{pia08}. Designations begin 
with `F6'.

\item  We had originally planned fields along the line betwen the SMC
 and LMC (the bridge): to test how the two galaxies interact near their
 extremities.  These were not executed in the end, but very similar
 data were obtained in an independent program by one of us \citep{har07}.

\end{enumerate}

\begin{figure}
\includegraphics[scale=0.70]{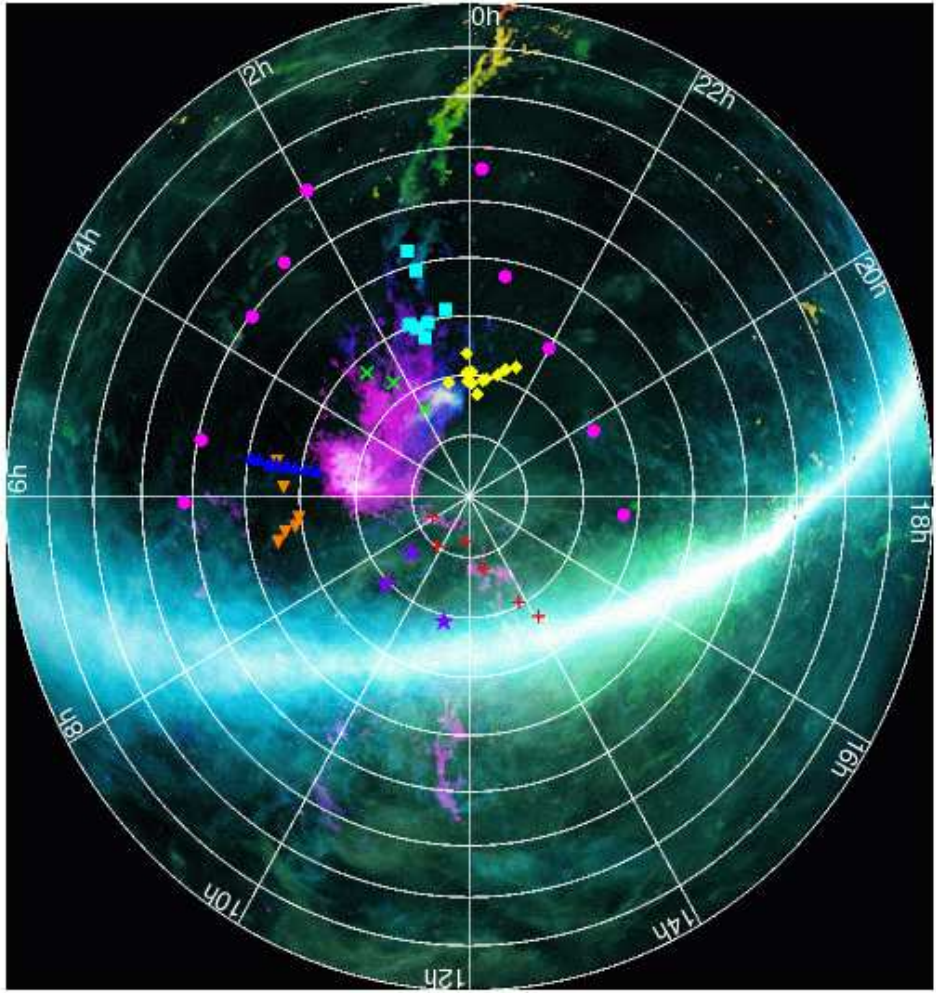}
\caption{ \scriptsize The figure shows the extended  region of sky surrounding the Magellanic Clouds, as seen
in neutral hydrogen emission.  The Galactic plane,  the location of the LMC and SMC, and both 
`leading' and `trailing' arms of the Magellanic Stream are clearly visible.  The positions of the fields
studied in this survey are shown by the over-plotted symbols. The main control fields are marked 
with purple filled circles, and the ancillary control fields whose designations begin with `F4C' are shown 
as purple asterisks. The fields marked in blue triangles are the ones going away due north from the LMC, roughly parallel to the Galactic plane.  The orange triangles show other fields that trace the 
extended LMC structure, with designations beginning with `F1'.  Fields tracing the trailing arm of the
Magellanic stream are shown as light blue squares: their designations begin with `F3'.  Red crosses 
mark fields along the leading arm of the stream, designations begin with `F4'.  
The yellow squares show fields with names beginning in `F5',  designed to trace the extended 
structure of the SMC. The blue crosses 
cut across the line of motion of the LMC, and designations begin with `F6'.  Image credit: S. Janowiecki and the 
Galactic All Sky Survey \citep{mcg09}.
\label{fig_fieldmap}
}
\end{figure}

The list of target fields actually observed is given in
Table~\ref{tab_targfields}, indicating field centers in both
equitorial and Galactic coordinates.  Of these, 45 have  designations 
beginning with the letter `F' and  were chosen
because their locations address one or more of the survey goals, as 
described above.   In
addition there are 10 fields beginning with `C', which
were designed as control fields for sampling background and Galactic
foreground contamination.  Our total region of interest spans a large
range in Galactic latitude and longitude, so these control fields
are necessary to trace changes in the Galactic foreground.  The
control fields were chosen to be far enough away from either Cloud, so
that the LMC/SMC complex cannot be expected to contribute to the star
counts at their locations. Figure~\ref{fig_fieldmap} shows the
locations of these fields in the spatial context of the LMC, SMC,
Galactic Plane, and HI in the Magellanic Stream, using results form the GASS 
survey \citep{mcg09}.  Three of the fields have
mixed designations (F4C1, F4C4, and F4C6): they are low Galactic
latitude, and are of interest in themselves, but they also were
intended to serve as comparison fields against  those that lie on the
`leading arm' of the Magellanic Stream.


\section{Data Processing}

\subsection{Processing of MOSAIC2 images}

The raw images from the MOSAIC2 were de-biased and flat-fielded using
dome flats (sky flats to correct for illumination did not improve overall flatness) 
and the standard IRAF\footnote{IRAF is distributed by the National Optical Astronomy 
Observatory, which is operated by the Association of Universities for Research in 
Astronomy (AURA) under cooperative agreement with the National Science Foundation}
tools in the MSCRED package.  The world coordinate systems (WCS) were 
derived and applied to each image, using
the routines MSCTPEAK or MSCSSETWCS plus MSCCMATCH, and matching
against the USNO-B1.0 astrometric catalog stars \citep{mon03}.
 With coordinates well established (typical rms errors of 0.2
arc-sec), the multi-extension FITS files containing the 8 image
sections from the 8 separate CCDs were then processed by the MSCIMAGE
routine in MSCRED to create a single image that is a gnomonic
projection. For any given target field, the position on the sky of one
specific CCD corner at the center of the mosaic for one of the long
$R$ exposures was defined as the tangent point for all images in all
bands of that field.  By this design, the projected images have pixels
of equal area. To zeroth order, they have the same pixel size as the
original image, but higher order terms correct the geometrical
distortions of the MOSAIC2 imager, and re-map the spherical
co-ordinates of the sky onto a plane.  Individual images of a given
field can then be cross-registered in position by simple translation
in the projected image plane, and co-averaged with the COMBINE routine in IRAF,
using position offsets driven by the fitted WCS.

The geometrical distortions in the MOSAIC camera means that pixels in different places in the 
field of view see slightly different solid angles on the sky.  The flat-field corrections, which 
have the net effect of mapping pixel value to surface-brightness,  induces 
a systematic error to aperture fluxes from point sources.  Re-mapping the image
to equal solid angle pixels by {\it resampling}, as opposed to conserving 
the pixel value sums in the flat-field corrected images, has the salubrious 
effect of restoring the image so that aperture fluxes are on an equal basis 
across the field.  It is important that the stellar PSFs be well sampled for this to work, 
a criterion that is always met for our data.  

Nonetheless, any re-binning of data generates 
correlation in the noise across neighboring pixels, while ignoring any pre-existing 
noise correlations that already exist in the input image.  Every point spread function fitting code 
known to us implicitly assumes that there is zero correlation of noise among neighboring pixels, 
so processes that increase correlation should be kept minimal.  Well sampled PSFs help 
to suppress errors from this source. A second effect is that that a spatial 
pattern in the noise is introduced due to `beating' of the old and new pixels, generating a 
Moire pattern in the noise.  To first order, both these effects result in increased noise in the 
photometry.  For an ensemble of objects taken over a significantly large patch of the image, the net scatter is increased, but there should not be systematic effects introduced for such an ensemble
(although for a single given object there may be subtle systematic effects involved, especially if 
images are under-sampled, which the MOSAIC2 images are not).   We have been diligent in watching for these effects, as discussed later in \S~\ref{iirspp}.

For each field, there are 3 final images in each of $C, R ~\&~ I$ to
cover the wide range of brightness:
\begin{enumerate}

\item A  {\bf deep} image obtained by a S/N-weighted co-average 
of {\bf all} images in that band. Many stars in this image  are saturated.

\item A {\bf medium} image obtained similarly by co-averaging the 
2 shorter exposures (for $R$ and $I$ the 50s and the 10s exposures,
and for $C$, the 300s and 60s exposures). Only the brightest stars are
saturated in this image.

\item A {\bf short} image, which is just the shortest image 
obtained (i.e. 10s exposure for $R$ and $I$, and 60s exposure in $C$).
All stars fainter than 12th mag should be unsaturated in these images.

\end{enumerate}

For $M$ and $DDO51$ images,  only  two final images are made in each band:

\begin{enumerate}

\item A {\bf deep} image which is the S/N-weighted co-average 
of {\bf all} images in that band, and

\item A {\bf short} image which is the co-average of the two short 
images (i.e. 30s exposures in $M$, and the 150s exposures in $DDO51$).

\end{enumerate}

The final images are masked as needed, so that any regions not exposed
(such as gaps between CCD chips) in all the component images are fully
suppressed for subsequent analysis.  All saturated pixels are assigned  a large
negative number value that the subsequent photometry programs interpret as 
missing data.  The FITS headers of the final
combined images were edited manually to carry the correct values of
GAIN (in electrons per ADU) and read-noise (in electrons), correctly
reflecting their propagation through the COMBINE processing.  Thus, in
the end, for each field, there are 13 final images.  Each image is
geometrically flat, and fitted with an accurate WCS.  They each carry
the correct values of gain and readnoise, as well as the correct
effective exposure time to which the co-averaging is scaled.
These images and photometry are being placed in the NOAO archive, as papers are published.

\subsection{Instrumental Photometry from  MOSAIC2 Images}
\label{sec_photometry}

Each of the 13 final images (as described above) of any given field
are then run through a  process constructed around a
variant (by one of us: Saha) of the DoPHOT photometry program
\citep{sch93}.

First an IDL based routine written by us (Brondel \& Saha) finds
the brighter objects, does a rudimentary rejection of galaxies and
cosmic-ray-like features by examining image roundness and sharpness.
It uses this preliminary set of what must be stellar objects, to
derive an analytic PSF in the form expected by DoPHOT.

Using the above derived initial PSF estimate, DoPHOT is run on the
image.  In addition to the list of objects and PSF fitted photometry,
this variant version also generates aperture magnitudes for a range of
aperture sizes of the bright high S/N stars, which are measured in
isolation, i.e. with all other objects subtracted.  There are two sets
of aperture sizes: set-A for aperture sizes from 2 to
16 pixel radius in steps of 2 pixels, and set-B for aperture radii
from 4 to 32 pixels in steps of 4 pixels.  For set-A, for each star to
which the procedure is applied, the sky subtracted is that scalar
value for which the dispersion in measured brightness for apertures
sizes 10, 12, 14 and 16 pixels is minimized, and for set-B the sky
value subtracted is similarly that for which the dispersion in
brightness for 20, 24, 28 and 32 pixels is minimized.  In effect this
procedure seeks that value of sky for which the growth curve is as
flat as possible for the outlying apertures of each set.  The
rationale for these two sets of apertures is explained below.

The set-B aperture magnitude at 20 pixel ($ \sim 5.4$ arc-sec)
radius, denoted by $m_{20}$ is the aperture to which we wish to refer
all measured magnitudes. For images with seeing $\leq 2.0$ arc-sec
FWHM, it is deemed to contain all of the incident light from a star,
except that which is scattered by the telescope and instrument optics.
Even though the seeing may vary from one image to another, all of the
seeing induced broadening is asserted to be within this
aperture. Another way of saying this is that the fraction of light from
a star which falls outside this aperture is from scattering, which
does not change (in any given passband) from exposure to
exposure. Thus, as long as the seeing is not larger than 2.0 arc-sec
FWHM, all stars in all exposures send the same fraction of their light
outside this aperture.  Thus $m_{20}$ can be used as an instrumental
magnitude, in the sense that it measures the same fraction of light
that reaches the telescope from a star for all stars in all exposures.
We must map the PSF fitted magnitudes (denoted by $m_{fit}$).  If the
PSF is constant within any given exposure, then all that is needed is
to calculate $ \langle m_{20} - m_{fit} \rangle $ using the brighter
stars, and apply this aperture-correction to the fitted magnitudes:
this is common practice for instruments where the PSFs are in fact
invariant over the (usually small) field of view.  A test for the
validity of a constant PSF is to look for position dependent trends in
$ m_{20} - m_{fit} $.  For our MOSAIC2 data, this test shows
significant trends, and a total scatter that in the worst situations 
can be as large as 0.2 mag!  Efforts to characterize the systematics
of this variation were thwarted by the fact that due to the large
aperture sizes, only a few objects have
high enough S/N measurements of $m_{20}$.  The set-A analog of
$m_{20}$ is $m_{10}$.  One can make much higher S/N measurements of
$m_{10}$ because of the smaller aperture size. Many more stars can
be measured with the required accuracy and subtraction of the fitted PSFs of 
neighboring objects has left fewer residues.  The aperture correction
systematics across the field of view (FOV) for any given exposure can
be far better traced using $ m_{10} - m_{fit} $ as compared to $
m_{20} - m_{fit} $.  However seeing changes from one exposure to
another can induce small systematic differences, so it may be too
small for use as an instrumental magnitude.  So the procedure is
broken into two parts: $ m_{10} - m_{fit} $ of a relatively large
number of stars is used to trace the aperture correction variations
across the FOV of a given image, and to apply suitable corrections
(details below), In another step, the value of

\begin{equation} 
 \Gamma = \langle m_{20} - m_{10} \rangle ~ \label{eq_Gamma}
 \end{equation} 

is evaluated from a few very high S/N stars, and applied as a further
correction, which finally refers all magnitudes to the instrumental
system of $m_{20}$.

Using $ \Delta m = m_{10} - m_{fit} $ to trace the aperture correction
systematics has proved to be very revealing:

\begin{enumerate}

\item 
The variations in $ \Delta m_{10}$ across the FOV are most acute when
the seeing is best. This is when the PSF variations across the field
are also the most prominent, so this confirms that the aperture
correction variations are induced by PSF variations.

\item
The variation of $\Delta m_{10}$ is smooth across any part of the FOV
covered by the same CCD chip, but there can be discrete jumps from one
chip to the next.  Fitting a single chip with a surface linear in $x$ and $y$
significantly reduces the scatter of $\Delta m_{10}$ within the area spanned by that
chip, but rarely eliminates it completely. 
\end{enumerate}

These results suggest that small mis-alignments between the chips
could be the source of much of the $\Delta m_{10}$ variation.  Each
chip is slightly non-orthogonal to the optical axis, which a linear
term in $x$ and $y$ corrects, and each chip has its own positional
offset from nominal along the optical axis, resulting in different
constant additional terms.  We also tried fitting a quadratic surface, 
and found that in all instances the scatter in the residuals reduces 
to levels consistent with the measurements errors.  
 Relatively large non-orthogonality in
position can  also result in quadratic terms, and is a possible explanation. 
However, further investigation reveals that the quadratic terms are highly 
correlated across the chips.  To understand this better, we 
examined the residuals after applying a linear surface correction to the 
individual chips, but studied them as a whole across the entire field of view.
We found a pattern in these residuals that is well fitted by a single
radially symmetric quadratic term.  Further, in the most pronounced
instances of $\Delta m_{10}$ variation, it is this component that
dominates.

\begin{figure}[htbp]
\includegraphics[scale=0.7]{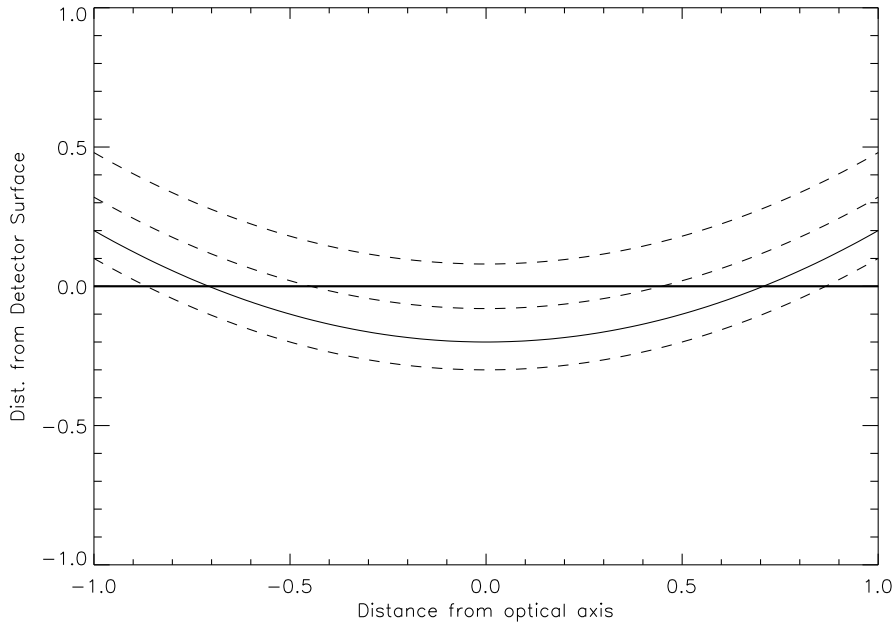}
\caption{\small The figure illustrates how variations in the PSF across the field of view (FOV) 
can arise because the focal surface is not an exact match to the
detector surface. The detector surface, here idealized as a perfect
plane, is viewed edge on: the horizontal line at ordinate zero
is the detector plane. The optical axis is vertical, and intersects
the detector surface at the center.  The surface of best focus,
imperfectly matched to the detector plane is shown as a curve. The
solid curve corresponds to an optimal compromise focus position, with
equal areas of the detector plane on either side of the focal surface.
It is clear that the farther away a given point on the detector is
from the focal surface, the more the PSF is degraded. Sub-optimal
focussing, where the position of the focal surface with respect to the
detector plane are shown by the dashed curves, makes the situation
worse.  If the detector surface is not a perfect plane orthogonal to
the optical axis, then that too will contribute to the run of PSF
variation across the FOV. In the case of MOSAIC2, in reality the
detector surface is in eight planar sections, each with its own
imperfection in alignment with respect to the ideal detector
plane. \label{fig_focdiag}}
\end{figure}

The origin of this radially symmetric variation in aperture correction
almost certainly lies in the interaction between the detector and
focal surfaces.  The true focal surface is a bowl, and the idealized
detector surface is flat.  The intersection of these two surfaces is
the locus of best focus, and clearly the PSF can be expected to vary
as the space between the two surfaces changes with position on the
FOV.  This is illustrated schematically in Fig.~\ref{fig_focdiag}. The
straight thick line at zero ordinate is the detector plane.  The ideal
focus is where the curved focal surface is set so that equal areas of
the detector plane lie on opposite sides of the surface: this is
represented by the full line that intersects the detector plane.  Less
optimal focus positions are shown by the various dashed lines, where
the focal surface is positioned non-optimally with respect to the
detector plane.  This graphically illustrates how differences in focal
placement drive quite different variations of the PSF with position on
the FOV.  Note that the aperture correction
variations are unique for each image, and must be evaluated
independently for each image on which photometry is being done.

At the end of our experimentation, it was determined that the most
robust constraints of the aperture correction variation as function of
position on the FOV are obtained by fitting the following elements
simultaneously:

\begin{enumerate}

\item
A quadratic surface symmetrical about the image center (assumed
optical axis intersection)

\item
A linear surface (plane), including offsets determined independently
for each section spanned by a different CCD chip in the FOV.

\end{enumerate}

Denote the final surface fit to $ \Delta m_{10} $ by $\Sigma_{10} (x,
y)$.  If $m^{i}_{fit}$ is the PSF fitted magnitude of the $i$th star
on the image at hand, and if it is located at position $(x,y)$, then
we can write:

\begin{equation}
 m^{i}_{inst}  = m^{i}_{fit} + \Sigma_{10}(x,y) + \Gamma ~,
 \label{eq_Minst}
\end{equation}

where $m^{i}_{inst}$is now the 10 pixel aperture equivalent magnitude
propagated from the PSF fitted magnitude, but on the system of a 20
pixel ($\sim 5.4$ arc-sec) radius aperture.

Routines to fit the surfaces were custom written in IDL by us (Saha \&
Brondel), and use weighting schemes that follow the error estimates
for each object as generated by DoPHOT.

Further calibration of the objects require establishing reference to
standard stars, and is discussed after we describe the processing of
the calibration images obtained with the 0.9m telescope.

\subsubsection{Implications of Image Re-Sampling on the PSF fit Photometry}
\label{iirspp}

The MOSAIC2 images are well sampled in all instances, even in the best
seeing we encountered.  In principle, a single resampling of the
images using sinc interpolation should not produce noticeable systematic
errors in the aperture referenced magnitudes derived as above. 
Nevertheless, in order to verify this empirically,
we ran several test comparisons, where we compared the photometry
performed on un-rebinned data on individual chips of single exposures
(corrected for pixel area variations derived from WCS fitting) against
that from the corresponding images as processed above.  The
comparisons are very satisfactory, with chi-square values (error
estimates from DoPHOT) that are significantly smaller than unity
(i.e. the differences are smaller than the the Poisson S/N errors).

\subsection{Processing of the 0.9m CFCCD images}
The raw images obtained with the CFCCD on the CTIO 0.9m telescope were
corrected for bias and flat-fielded (using a combination of dome and twilight
flats taken on the same night), using the QUADRED package in IRAF.
During each observing run, daytime observation of the dome flat-field
source were used to create a shutter timing/shading correction: this
is a correction {\it image}, that is used to correct both, the short
exposure flat-fields (especially of the twilight sky), and all target
exposures so that the intensity at every pixel of the image is scaled
from how long that pixel was really exposed to the nominal exposure
time for that image.


The data were taken so there are always a pair of images in each of
the four bands ($R, I, C, M$).  However, the images were not combined:
photometry was performed independently on each exposure, and then
merged, using error estimate weighted averaging, and propagating the
resulting uncertainties.  For this reason, there was no compelling
reason to resample the images. Any variations of pixel size as
projected on the sky were corrected using a pixel-area correction
derived from the WCS solutions and the images scaled accordingly.

\subsection{Instrumental Photometry from CFCCD Images}

The photometry process was like the one described
above for the MOSAIC2 images, except for a few differences in
parameters resulting from differences in pixel scale, and the fact
that there is only one CCD chip, and  thus only one correction for the 
tilt between the detector and focal surfaces.
Accordingly, the set-A aperture sizes in
pixels for the CFCCD data were identical to those for the MOSAIC2
data.  However, here the equivalent $m_{10}$ corresponds to $\sim 4$
arc-sec.  The set-B apertures for CFCCD were 3, 6, 9, 12, 15, 18, 21,
\& 24 pixels in radius.  In lieu of $m_{20}$ used for the larger
aperture magnitude for MOSAIC2 data, for CFCCD data we use $m_{15}$,
which corresponds to $\sim 6.0$ arc-sec, which on the sky is only 10\%
larger than used for MOSAIC2.  In the correction from $m_{fit}$ to
$m_{10}$, $\Delta m_{10}$ was fit to a quadratic surface symmetrical
about the center of the FOV. Again, the complications that come with
having multiple CCD chips do not appear here.

The final instrumental magnitude $m_{inst}$ for the CFCCD, is then the 
magnitude propagated from the PSF fitted magnitude, but on
the system of a 15 pixel ($\sim 6.0$ arc-sec) radius aperture. 
This instrumental magnitude is uncorrected for
extinction from the terrestrial atmosphere, 
as is the case for the  MOSAIC2 instrumental mags.
.

\subsection{Calibrated Photometry from the CFCCD Data}
\label{sec_calCFCCD}

On each night of observing with the CFCCD, target object fields were
interspersed with standard star fields from \citet{lan83,lan92}.  
These references furnish $R$ and $I$ values for
several stars per field.  \citet{gei96} provides $C$ and $M$ standards
for select stars in SA92,
SA98, SA101, SA107 and SA114, NGC~3680 and around PG0231+051. In an
unpublished work, Saha has used observations from the WIYN telescope
to establish $C$ and $M$ magnitudes of select stars in Landolt fields
in SA92, SA98, SA110, Ru149 and M~15.  Cross-comparison with photometry in 
\citet{gei96}, and with additional unpublished  photometry of  the globular cluster M15 
kindly provided to us by
 \citet{gei05} were used to achieve this calibration. These  $C$ and $M$ standards in the 
 Landolt fields are 
established to be internally consistent to $\sim$ 1\%: for
instance the Saha set calibrated using only M15 inter-comparison
predicts magnitudes of stars in SA114 that agree with the values of
\citet{gei96} to better than 0.015 mag in the mean over the entire
color range $ 0 < C-R < 5 $.  The observations of these chosen Landolt
fields thus provide standard stars for both Landolt $R$ \& $I$, as
well as for Washington system $C$ and $M$ bands.

For any given target, the instrumental magnitudes in the four bands
for each target are collated. The object list from the $R$ image is
treated as the master. Objects from the other three filters are
matched to that, on the basis of position on the sky, with typically a
1 arc-second match tolerance.  Thus the collated list always has a
measurement in $R$, but for faint objects, there may be drop-outs in
one or more of the other bands.

\begin{figure}[htbp]
\begin{center}
\includegraphics[scale=0.6]{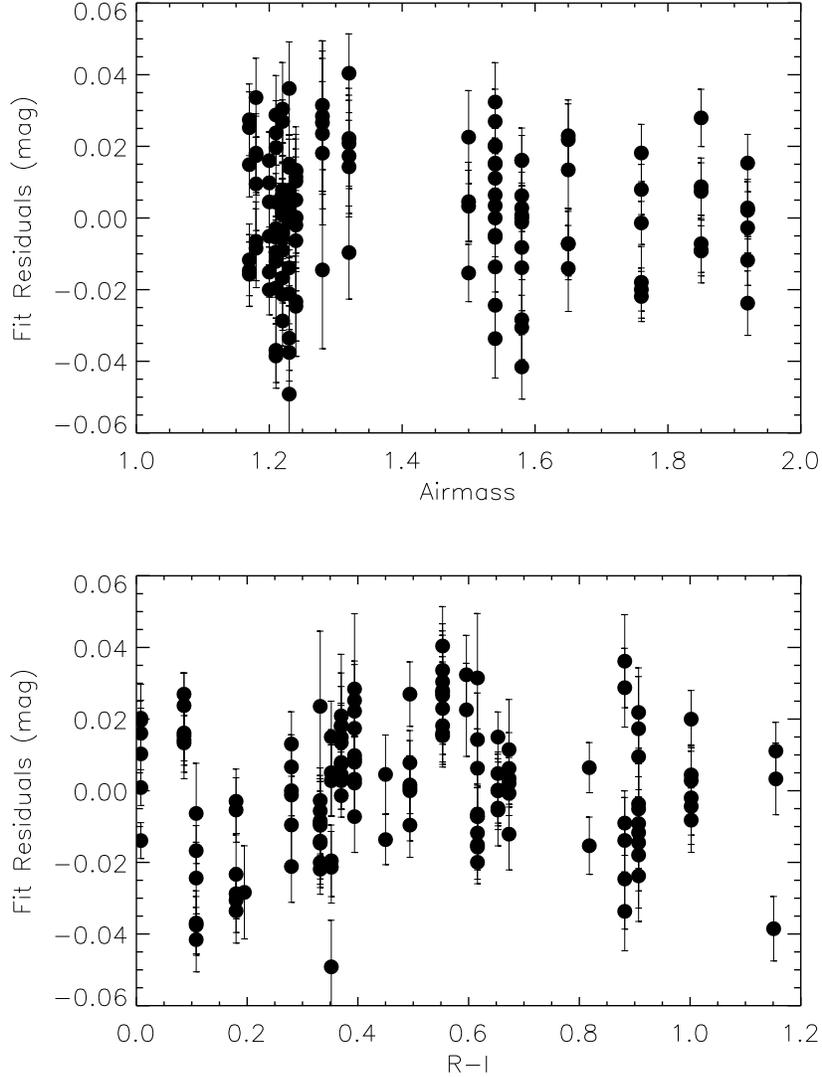}
\caption{The plots show the fit residuals in the $R$ band for the night of 2007 Oct 12, from CFCCD observations of standard stars.
The upper panel shows the residuals against observation airmass, and the lower panel shows the same against the $R-I$ value of the respective standard star.  There were 23 different standard stars, each observed several times. The error estimates for each measurement is shown: the solution used the inverse square of the error bars as weights.  See Table~\ref{tab_ntcoeffs} for the solutions for this night. 
\label{fig_r_residuals}
}
\end{center}
\end{figure}

Consider the $R$ band as an example. Denote the instrumental
magnitudes by $R^\prime$.  First, $R^\prime$ for the standard stars are
compared against their true values $R$. Allowing for variations due to
extinction and color dependence, we solve for:

\begin{equation}
R = R^\prime  + \alpha + \beta X + \gamma  COLOR
\label{eq_nightcoeffs}
\end{equation}
where, $R$ is the true magnitude, $X$ the airmass at which the
observation was made, and $COLOR$ is a suitable quantity, e.g. $R-I$,
and $\alpha, ~\beta$ and $\gamma$ are coefficients that are solved
for, using many measurements of several standard stars described above
( $> 100$ measures in each of $R$ and $I$, and $> 60$ measurements in
each of $C$ and $M$ per night) spanning a wide range of colors, and
airmass range from $1.2 < X < 2.1$.  All fitting uses individual error
estimates propagated from DoPHOT for
weighting. Table~\ref{tab_ntcoeffs} lists the coefficients from the
solution for the night of 2007 Oct 12, as an example of the values and
the residuals.  Fig~\ref{fig_r_residuals} shows residuals from the
same night plotted against both airmass and color of the
star. The rms scatter is less than 0.02 mag, typical of nights that
were {\it a posteriori} considered photometric. Nights when any of the
four pass-bands show rms scatter exceeding 0.03 mag were discarded,
since conditions  may be suspect.

Once the coefficients in eq.~\ref{eq_nightcoeffs} were evaluated, and
the fit residuals found to be satisfactory, the 0.9m telescope instrumental
magnitudes of the target object fields were transformed to true
magnitudes on the system of Landolt or of Washington (as appropriate
for that band), by inverting eq.~\ref{eq_nightcoeffs} and using the
now known values for $\alpha, ~\beta, ~and ~\gamma$.

\subsection{Transfer of Calibrated Magnitudes to the MOSAIC2 results}
\label{sec_calxfer}

Tranferring the calibration to the MOSAIC2 data  involves several steps.  
First the instrumental
magnitudes from the deep, (medium,) and shallow images from MOSAIC2
for any pass-band are merged.  Stars in common are recognized, and any
offsets in instrumental magnitudes are adjusted. Such offsets can
occur because of observations at different airmass, or due to throughput differences 
on different observing runs. For a
given band, a single list is created, which contains objects from all
the final images in that band.  Where an object occurs on two or more
of the lists, the weighted average is taken (weighted by inverse
variance) using individual values for each star.  The propagated error
is also the weighted error derived from the component error values.
The combined list for the $R$ band serves as the master list, and the
final lists from the other four bands for MOSAIC2 ($C$, $I$, $M$ and
$DDO51$) are matched to the master list.  A combined MOSAIC2 list is created,
which now has an entry for each star detected in $R$, with
instrumental magnitude and error values in each band for which a
measurement is available.  By construction, there is always a
measurement available for $R$.

\begin{figure}[htbp]
\includegraphics[scale=0.7]{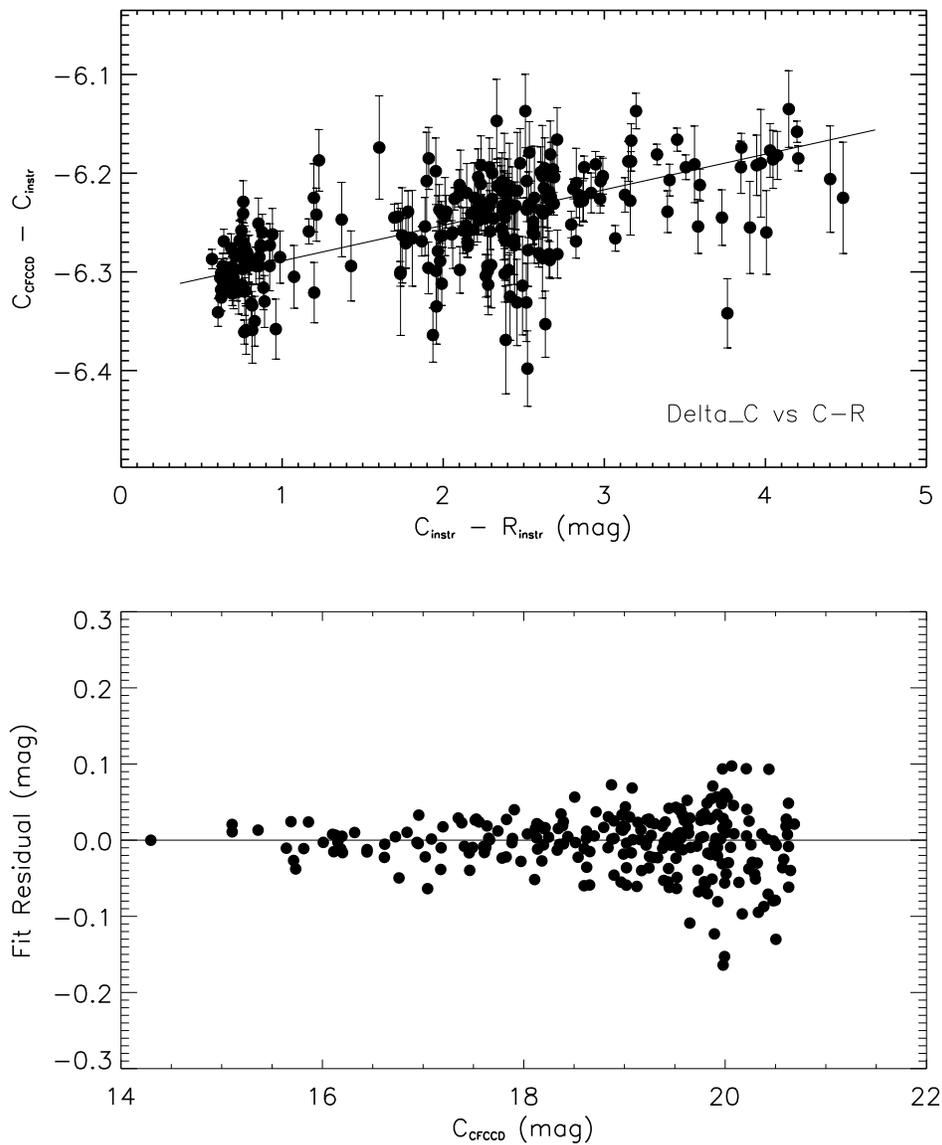}
\caption{\small This figure illustrates the calibration of MOSAIC2 magnitudes as described 
in \S\ref{sec_calxfer}. The upper panel shows the color dependent mapping from instrumental magnitudes in the $C$ band measured with MOSAIC2 (denoted by $C_{instr}$) against $C$ magnitudes on the 
Washington system established for CFCCD observations (denoted by $C_{CFCCD}$) as described in 
\S\ref{sec_calCFCCD}.  The lower panel shows the fits residuals as a function of true $C$ magnitudes.
This example shown in for the field F601. The rms scatter in the fit, which includes objects spanning 6 mags
in brightness, is 0.039 mag, with a weighted uncertainty in the mean of only 0.0023 mag.
\label{fig_calxfer}
}
\end{figure}

The instrumental magnitudes from any given image derived from MOSAIC2
observations are expected to differ from true magnitudes by a
zero-point adjustment, and a first order color term.  Atmospheric extinction suffered by the MOSAIC2 images is
subsumed in these two terms.  Accordingly, we can write (for example):

\begin{equation}
R ~~=~~ R^{\prime} +   {\cal A}  ~+~  {\cal B}  (R^\prime - I^\prime) 
\label{eq_photxfer}
\end{equation}

where $R^\prime$ and $I^\prime$ are instrumental mags for a given star
from MOSAIC2, and $R$ and $I$ are the true values.  The coefficients
$\cal A$ and $\cal B$ can thus be derived.  Any pass-band and
instrumental color can be used in the form shown above, each with
their respective coefficients.  In practice, given that we use only
those stars for which measurements exist in $R$, for any band $X$,
where $X \neq R$, the instrumental color color $(R^\prime - X^\prime)$
is always available.  For $R$ itself, we solve for alternate colors:
$R^\prime - I^\prime$, as well as $C^\prime - R^\prime$, which covers
objects too blue to be detected in $I$, as well as those too red to be
detected in $C$.  Typical residuals for a fit to equation \ref{eq_photxfer}
are shown in Fig.~\ref{fig_calxfer}.
Once the coefficients are derived from the stars in
common to MOSAIC2 and CFCCD, the values can be applied to all stars in
the MOSAIC2 list for $R$, $I$, $C$, and $M$.  These are the final
calibrated magnitudes on the Landolt system for $R$ and $I$, and on
the Washington systems for $C$ and $M$.

Since $DDO51$ magnitudes are not obtained with the CFCCD (it is
unnecessary, and it would take too long an exposure to get sufficient
numbers of stars to match against MOSAIC2), the $DDO51$ measurements
at this point are uncalibrated.  We force an artificial calibration
using the precept that the metal absorption features in the $DDO51$
band do not form sufficiently in stars that are hot, or more specifically,  that 
differences in relative transmission through the $DDO51$ and $M$ 
passbands for such stars are not significant.
Since the $DDO51$
pass-band lies in the middle of the $M$ pass-band, we force the
$DDO51$ band to equal the $M$ band mag on average, for all definite 
stars brighter than $I = 20.0$ with estimated measurement errors in the 
$M$ and $DDO51$ bands  less than 0.05 mag and estimated $I$ error less than 0.1 mag
that have $M-I < 1.0$.  
We denote this ersatz $DDO51$ {\it magnitude value} by $DDO51s$ 
(though we continue to refer
to the passband by $DDO51$). 

\subsubsection{Consequence of not using independent color terms for each CCD}

The procedure above makes the tacit assumption that the color
responses of the 8 CCDs of MOSAIC2 are identical, so that a single
pair of coefficients $ \cal A$ and $\cal B$ can be used in
equation~\ref{eq_photxfer}.  While the flat-field normalization does a
zeroth order balancing of the responses, color terms remain because
the color of the flat-field is not the same as the color of a star,
and stars themselves span a large range of colors.  The overlap of the
2 CFCCD pointings over each MOSAIC2 field allows for common stars to
be found on all 8 CCDs of MOSAIC2, so in principle one can solve for 8
different values of $\cal A$ and $\cal B$, one pair for each CCD.  The
procedure would thus be quite straightforward, but would {\it
require} that observations exist with both pointings of the CFCCD.
Also, in the fields with higher Galactic latitude, the number of high
S/N stars measured in the overlap area with any one MOSAIC2 chip can
get quite small, thus incurring larger uncertainties due to random
errors of measurement.  Another strategy could be to document the
color response difference for each chip with respect to the mean
obtained from comparing 4 or 8 of the CCDs together. 

To evaluate the chip to chip variations, on a photometric night (otherwise unusable because of 
 poor seeing) we obtained observations of a
standard field, placing the same stars in turn on all eight MOSAIC2 CCDs.
Using the $R$ band as an example, and using all available measurements
on all chips, we first estimate the effects of extinction to zeroth
order, and solve equation~\ref{eq_nightcoeffs}.  This forces an
initial solution assuming no color dependence variations from one CCD
to another.  We retain the value of $\beta$ (airmass dependence) from
this solution.  Next, we force the above derived values of $\beta$,
and construct the extinction independent (to first order) instrumental
mags for the $R$, $I$, and $C$ bands as follows:

\begin{eqnarray}
R^{\prime\prime} = R^{\prime} - \beta_{R}  X \\
I^{\prime\prime} = I^{\prime} - \beta_{I}  X  \\
C^{\prime\prime} = C^{\prime} - \beta_{C}  X 
\label{eq_Xindep}
\end{eqnarray}

Using these airmass dependence corrected instrumental mags, we then
solve the following equations {\it independently} for each CCD:

\begin{eqnarray}
R ~~= ~~R^{\prime\prime}  ~+~  {\cal A}^{\prime}_{R}  ~+~ {\cal B}^{\prime}_{R} . (R^{\prime\prime} - I^{\prime\prime}) \\
I  ~~=~~ I^{\prime\prime}  ~+~  {\cal A}^{\prime}_{I}   ~+~ {\cal B}^{\prime}_{I} . (R^{\prime\prime} - I^{\prime\prime}) \\
C ~~=~~ C^{\prime\prime} ~+~ {\cal A}^{\prime}_{C}  ~+~ {\cal B}^{\prime}_{C}.(C^{\prime\prime} - R^{\prime\prime})
\label{eq_coldep}
\end{eqnarray}

Table~\ref{tab_colresponse} shows the values of ${\cal A}^\prime$'s
and ${\cal B}^\prime$'s for each of the three above passbands for each
of the eight CCDs, as measured on the night of 2005 Oct 3 (UT).  The
field contains 6 Landolt standards in SA 92: star numbers 245, 248,
249, 250, 252 and 253.  These stars span a color range in $B-V$ from
0.5 to 1.4, which covers the color range of interest for this survey.
Obviously, objects outside this range are interesting, but photometric
accuracy demands for the analysis of ages and metallicities from Hess
diagrams are well covered by this color range.
Table~\ref{tab_colresponse} shows that within this color range, the
rms scatter in recovering the standard star photometry is between 0.01
to 0.02 mags when analysis is done independently within each CCD.  If
a common solution is used with stars in all CCDs, the scatter
increases marginally, to about 0.02 mag.  These results are also
consistent with independent analysis in the $C$, $M$, $T1$ and $T2$
bands by one of us \citep{ols09} using separate and independent data contemporaneous
with this survey.  We estimate, using the data in
Table~\ref{tab_colresponse} , that at $B-V = 0.0$, we could make a
systematic error in color by at most 0.02, 0.03, and .04 mag in $R$,
$I$, and $C$ respectively, by ignoring color response variations from
chip to chip. At $B-V = 1.5$, the errors can be as large as 0.04,
0.05, and 0.04 mag in $R$, $I$, and $C$ respectively.  These are extreme cases, and as
shown above, for the issues we seek to address, the rms errors
incurred of 0.01 to 0.02 mag are no larger than other sources of
error.  We have therefore chosen the robustness of a single color term
and zero-point for all chips, over the difficulty of accurately
pinning down the exact color-terms, since the scientific return for
doing so would be marginal at best.

\section{Color-Magnitude Diagrams of Fields along a northern extension from the LMC}
\label{sec_cmds}

In this paper we present the photometry for fields along a line going
due north from the LMC bar.  These 8 fields (Table~\ref{tab_targfields}) are F7N, F9N, F11N, F12p5N, F14N,
F123N, F122N and F121N.\footnote{The combined images and calibrated photometry for these fields are being made 
available in the NOAO survey program archives.} Their respective distances from the LMC center
(on the sky) range from  $7^\circ$ to  $19^\circ$.  They all lie at
a Galactic latitude $ b \approx -34^\circ$.  Their individual distances from the LMC center 
are listed in Table~\ref{tab_starcounts}.  In addition we consider the two fields 
F111 and F113, which flank the line traced by the above fields at LMC-centric distances 
between $12^\circ$ and $14^\circ$, which is where we originally expected 
to see a pile up of tidal debris, based on the tidal radius estimate by \citet{wein00}
done before the new and improved proper motion of the LMC was known (although 
we should point out that his estimate does not demand the orbit).

\subsection{Comparison with Isochrones}  
\label{sec_cmd-isoc}

\begin{figure}[htbp]
\includegraphics[scale=0.6]{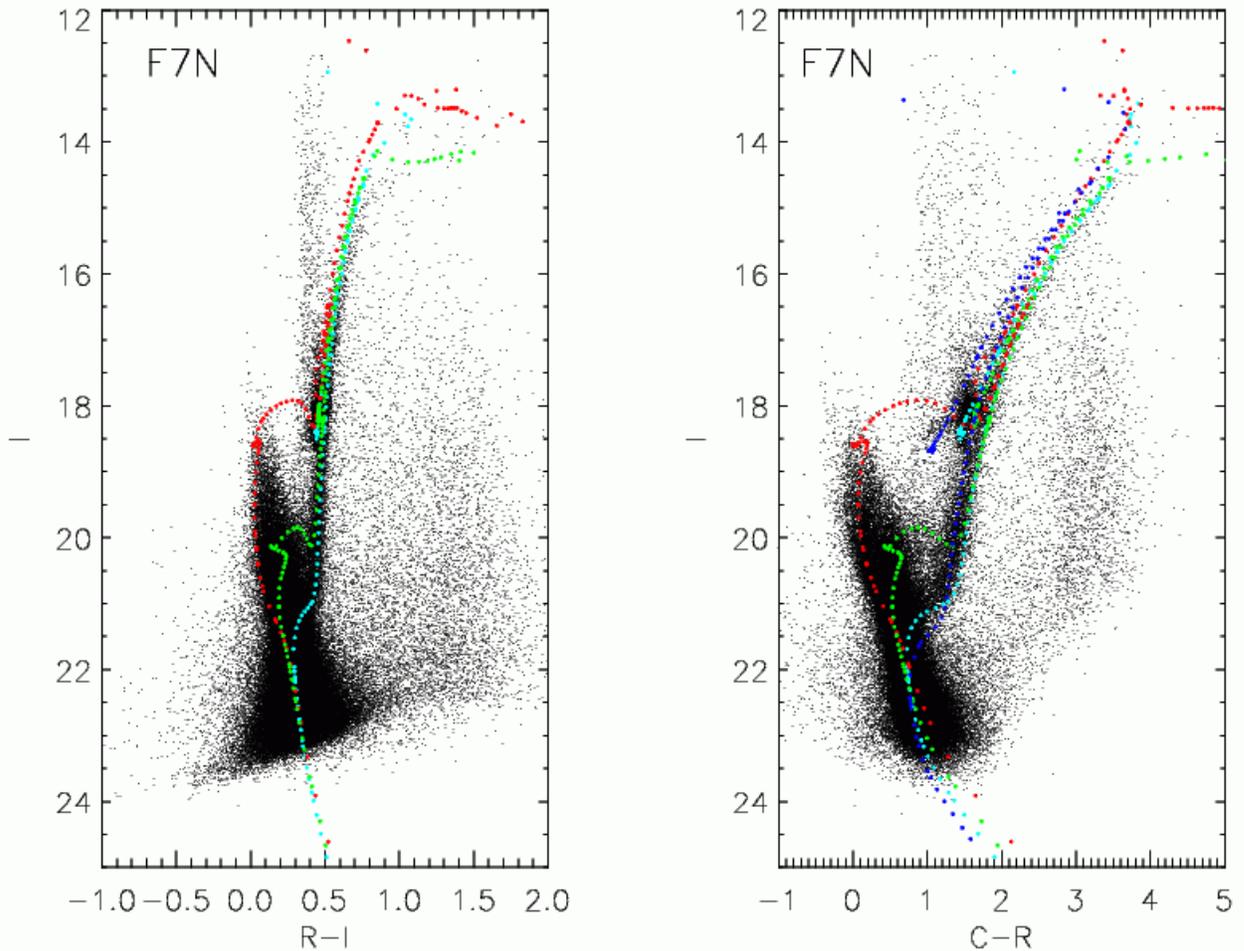}
\caption{\small The color-magnitude diagrams of the F7N field are shown: the left panel shows $R-I$ vs $I$, and the right panel shows $C-R$ vs $I$.  Only objects whose PSF profiles are unambiguously consistent with those of a stellar PSF are shown.  Several isochrones from Marigo et al. (2008) are overplotted, assuming a distance modulus of 18.55, and reddening $E(B-V) = 0.05$: the isochrone in dark blue
is for $Z=.001$ and $ \log t = 10.15 $, the light blue is for $Z=0.002$ and $\log t = 9.9$, the green if for 
$Z=.004$ and $\log t = 9.3$, and red for $Z = .008$ and $\log t = 8.7$.  These are not rigorous fits
to the data, but serve to illustrate the range of ages present, and that progressively younger stars are also progressively metal rich.  We also note the foreground wall of disk, thick-disk and halo turn-off stars at $R-I \sim 0.25$ and the pile-up of cool stars at $C-R \sim 3$.  \label{fig_F7Nisochrone}
}
\end{figure}

Photometry in $CRI$ for the field F7N, which is the closest to the LMC
center, is shown as two CMD's (one with $R-I$ as color, and the other
with $C-R$ as color) in Fig.~\ref{fig_F7Nisochrone}.  Only objects
that are classified definitely to be stars are plotted, which means
that features that extend a magnitude fainter are not shown because
there is not adequate S/N for them to be unambiguously distinguished as stars 
or as background compact galaxies.  The CMD's show a well defined giant
branch and red clump/horizontal branch.  The turn-off stars span a range of 
brightness: from about $I \sim21.0$ at the faint end to those still on the main 
sequence extending as bright as $I \sim 18$

Select isochrones from \citet{Mar08} (obtained from the web-site {\it
http://stev.oapd.inaf.it/cgi-bin/cmd}) are over-plotted in
Fig~\ref{fig_F7Nisochrone}, and annotated in the figure caption.
These tracks are `fit' by eye, and are no substitute for a rigorous
analysis of the Hess diagram, which will be the subject of a future
paper.  However, even the relatively rudimentary exercise of producing
these figures has been revealing.  The `fits' are constrained by having to accommodate the width of
the main-sequence (MS), the width of the sub-giant branch (SGB), and
the colors of the brighter red giant branch (RGB).  The dark blue
isochrone (shown only in the right hand panel of
Fig~\ref{fig_F7Nisochrone}) shows that 14~Gyr metal poor stars are
rare, for neither the SGB nor the upper RGB are fit by this isochrone.
But there must be some stars, perhaps 10~Gyr or older, to explain the
blue extension of the red clump into a horizontal branch, as seen in
the $C-R$ vs. $I$ CMD.  It is not possible to measure the star
formation rates without a full quantitative Hess diagram analysis (planned 
in the near future) but the isochrone fits indicate that stars older than
8~Gyr are rare.  Uncertainties of order 0.1 mag in
the distance modulus, or 0.05 mag in $E(B-V)$ reddening affect the
finer details, but not the overall conclusions.  The isochrones
`allowed' by the CMD constrain what age-metal combinations are
permissible: the oldest stars have $Z \sim 0.001$, and the youngest
about $Z \sim 0.008$.  This corroborates the finding by \citet{Gal08}
in their study of fields from $2.3^\circ$ to $7.1^\circ$ from the
center of the LMC, that younger stars must be progressively metal
rich. 

Predictably enough, it is the CMD in $C-R$ which
primarily drives and constrains the isochrone comparisons, and demands
that metals increase for younger stars. This justifies adding the
observationally expensive $C$ band data, which allows purchase on the
metallicity.  Changing the metallicity by a factor of 2 at a given age quite 
dramatically degrades agreement of data with isochrone; changing age by 
20\% at fixed metallicity also does the same.
 It is additionally satisfying that the chosen isochrones
cover {\it both} CMDs, i.e. in $R-I$ as well as in $C-R$, which lends
confidence and credibility to the outcome.  A future detailed Hess
diagram analysis will bring to bear all of the constraints on ages and
metallicities inherent in the $C,M,R,I$ bands, to provide a well
constrained star formation history.

\begin{figure}[htbp]
\includegraphics[scale=0.7]{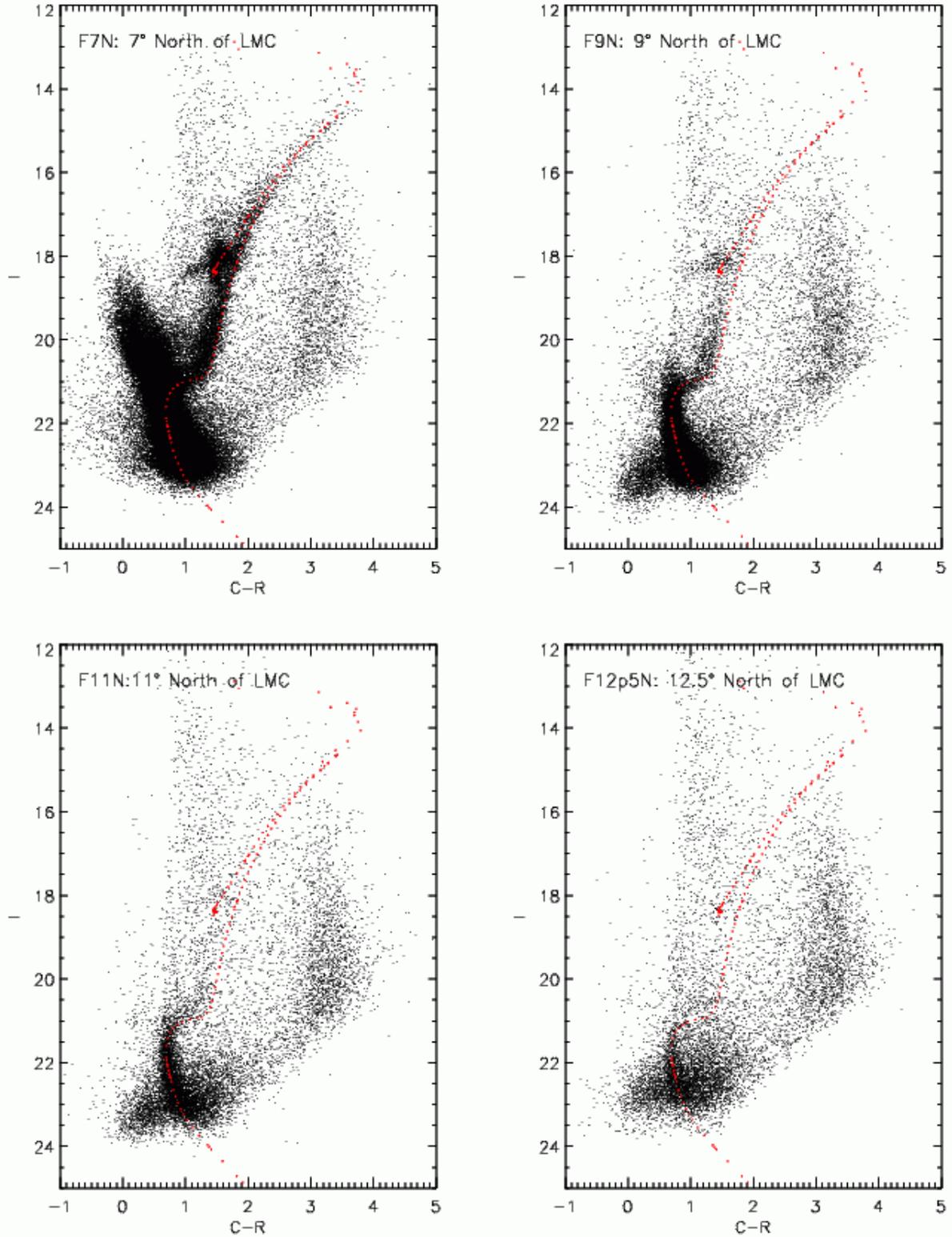}
\caption{\scriptsize The color-magnitude diagrams in $C-R$ vs $I$ are shown for fields progressively more distant
(from $7^\circ$ to $12.5^\circ$) along a line due North from the center of the LMC.  All objects whose profiles are consistent with those of
a stellar PSF are shown.  The majority  of objects  fainter than $I=22.5$ and with colors 
ranging from $0 < C-R < 1.5$, especially in the more distant fields, are background 
galaxies, unresolved in the seeing limited images.  The isochrone for $Z=0.002$ and $\log t = 9.9$ (8~Gyr) is over-plotted.  See \S~\ref{sec_cmd-isoc} for interpretation.
\label{fig_cmdprog1}
}
\end{figure}

\begin{figure}[htbp]
\includegraphics[scale=0.7]{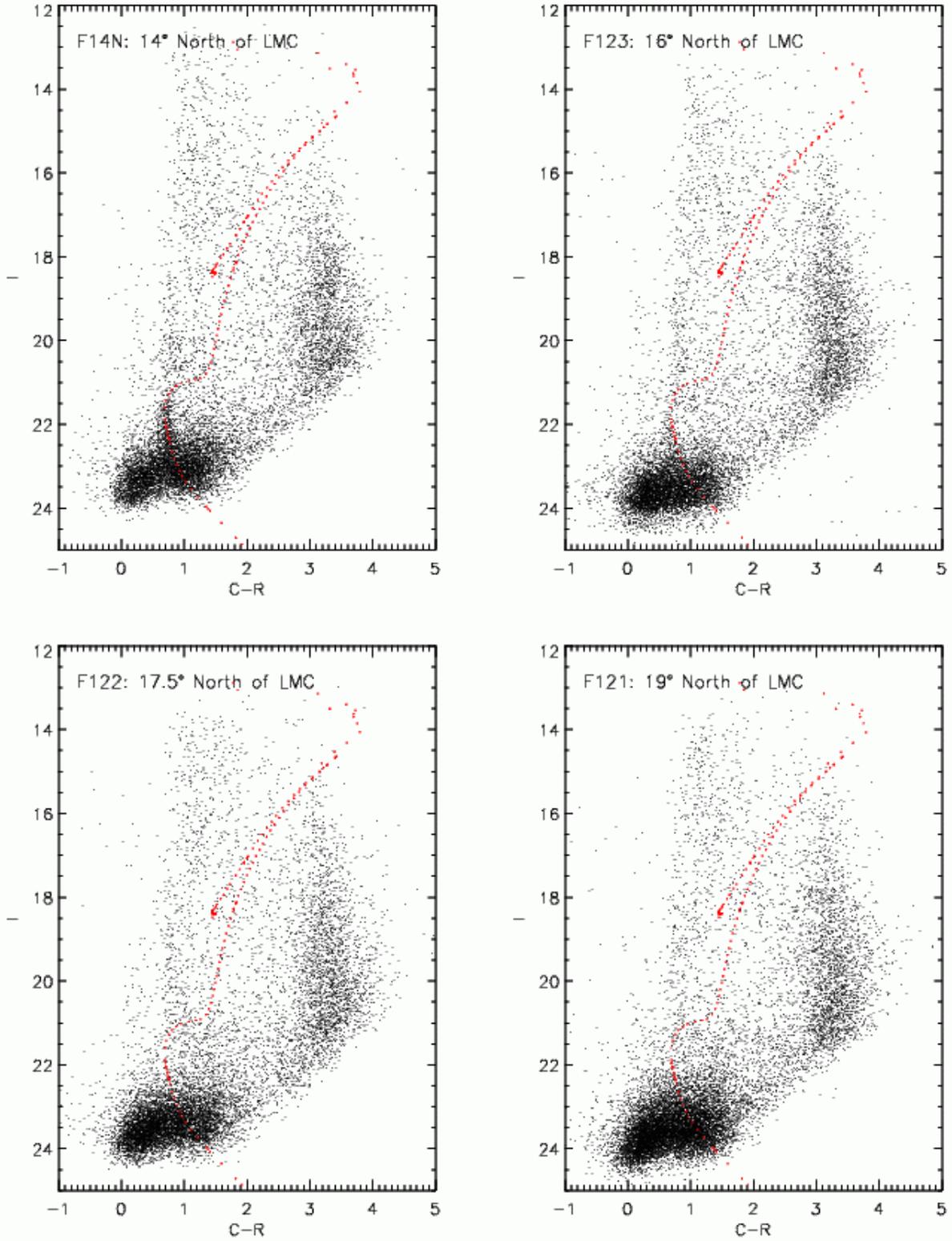}
\caption{\small Same as Fig~\ref{fig_cmdprog1}, but for fields at distances from $14^\circ$ to $19^\circ$ from the LMC center. \label{fig_cmdprog2} 
}
\end{figure}

\begin{figure}[htbp]
\includegraphics[scale=0.7]{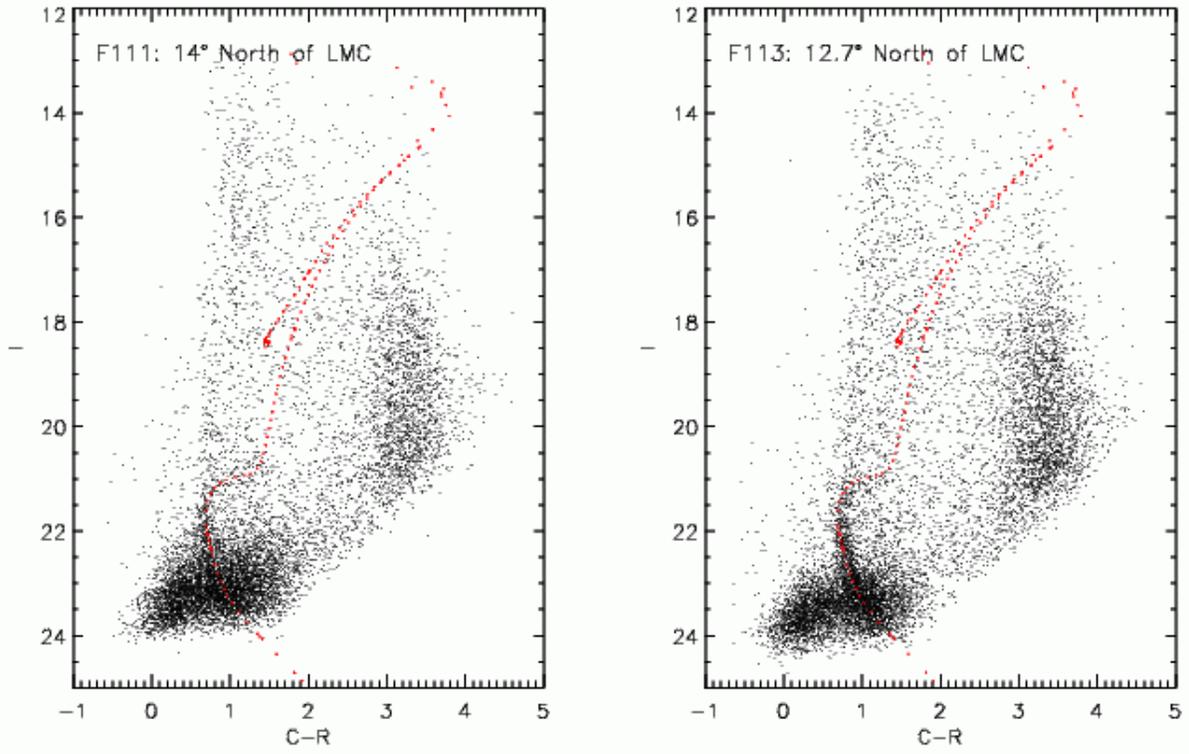}
\caption{\small Same as Fig~\ref{fig_cmdprog1}, but for the flanking fields fields between $12^\circ$ to $14^\circ$ from the LMC center. \label{fig_cmdprog3} 
}
\end{figure}

The CMDs in $C-R$ vs $I$ are shown for the fields with progressively
increasing distances, in Fig.~\ref{fig_cmdprog1} and
Fig.~\ref{fig_cmdprog2}.  The isochrone for $Z=0.002$ and $\log t =
9.9$ (8~Gyrs) at a distance modulus of 18.55 and reddened by $E(B-V) =
0.05$ is over-plotted on all 8 CMDs. The CMDs from the flanking fields 
F111, F113 are shown in Fig.~\ref{fig_cmdprog3}.

A smooth progression in the features of the CMD is apparent.  Going
from $7^\circ$ through $9^\circ$ to $11^\circ$, we see not only a
decline in the total number of stars present, but also a steady
erosion of the younger stars relative to the older ones, as seen from
the sharp decline in the number of blue MS stars (above the oldest
turn-off).  By the time we are $11^\circ$ out, the young stars are all
essentially gone.  The overall decline in the number of all stars also
causes the RGB to vanish against the `backdrop' of foreground stars
from the Galaxy: it is clearly delineated in F7N, still quite visible
in F9N, but not discernible on its own in the CMD of F11N.  In contrast, the MS
stars clearly continue to stand out prominently.

This is the expected validation of one of the basic precepts of this
survey: that of reaching the MS stars below the oldest turn-off, and
using them as tracers of extended structure.  Anticipating the counts
of stars in the MS described later in this section, we derive the
equivalent surface brightness in $I$ from the stars in the MS at the
location of F11N to be $\sim 30.5~ {\rm mag~arcsec}^{-2}$.  This is
several hundred times fainter than the sky brightness.  Surface
brightness measurements in galaxies farther away would not reveal the
equivalent sructure, and as evidenced here, RGB stars are too sparse
(nearly 100 times sparser than the MS stars) to be useful for tracing
kilo-parsec scale structure.  However the MS stars allow us to push
on.  Proceeding through the CMDs of fields even farther out (see
Fig~\ref{fig_cmdprog2}), the MS feature persists through the field at
$14^\circ$ (F14N) and is visible even in the $16^\circ$ field (F123).
In the two outermost fields (F122 at $17.5^\circ$ and F121 at
$19^\circ$), the last vestiges of the MS disappear into the ambient
stars in the line of sight.  In F14N, the equivalent surface
brightness in $I$ of the MS stars is $\sim 32.3$, and in F123 it is
$\sim 34.8 ~ {\rm mag~arcsec}^{-2}$.  Real surface brightness
detections of structure in any galaxy at such levels is not possible
with any foreseeable instrumentation, whether from the ground, or from
space, since these are $10^{4}$ times or more fainter than the ambient sky
brightness.  If the LMC were twice as far away than it is, the MS even
in the F11N field would disappear into the cloud of unresolved
galaxies for ground based observations, though it would remain
accessible to space based imaging.  The LMC (and the SMC) thus
provides us with an unprecedented opportunity to probe structure that
is bound to a parent galaxy, using MS star counts.

\subsection{Identifying giants using the $DDO51$ photometry}
\label{sec_ddo51giants}

The $DDO51$ passband, originally defined by \citet{cla79}, and introduced into the Washington 
system by \citet{gei84}, admits a narrow ($\sim 100$ \AA ~wide) 
part of the spectrum centered near $5150$ \AA, and includes 
the MgI triplet and bands of MgH, which are sensitive to surface gravity
(in addition to temperature and abundance).   These spectral features have been widely used
to separate G and K giants from dwarfs in the same temperature range from low dispersion
spectra. The $DDO51$ passband is able to do the same from appropriate photometric data.
The strength of the absorption features is measured by the index $M - DDO51s$, since 
$DDO51$  is conveniently situated in the middle of the $M$ passband. 
A second index must be used to track and de-trend the effects of temperature.
Details on implementing this technique are given by \citet{maj00} . They used $M-T2$ (Washington system) as the temperature index.  Their Fig.~4 demonstrates the 
sensitivity of the method for stars with colors redward of the turn-off for old stars.

An elaborate critical discussion of this technique, replete with caveats and limitations, 
is given in   \cite{mor01}.  Specifically they warn that errors in both $M-T2$, and $M-DDO51$
need to be held within a few hundredths of a magnitude to avoid specious detections of giants
because of photometric error driven contamination  from the much more numerous main sequence stars.

Our implementation has a couple of subtle differences from that of
both \citet{maj00} and \citet{mor01}.  We choose $M-I$ as the
temperature index, noting that $I$ is very close to the $T2$ band.
Also, instead of $DDO51$ magnitudes defined by \citet{gei84}, we use
$DDO51s$, as defined in \S~\ref{sec_calxfer}.  We define our procedure
empirically, using the photometric data for the field F7N, where the
giants are numerous, and easily seen in the CMDs.

\begin{figure}[htbp]
\includegraphics[scale=0.6, angle=90]{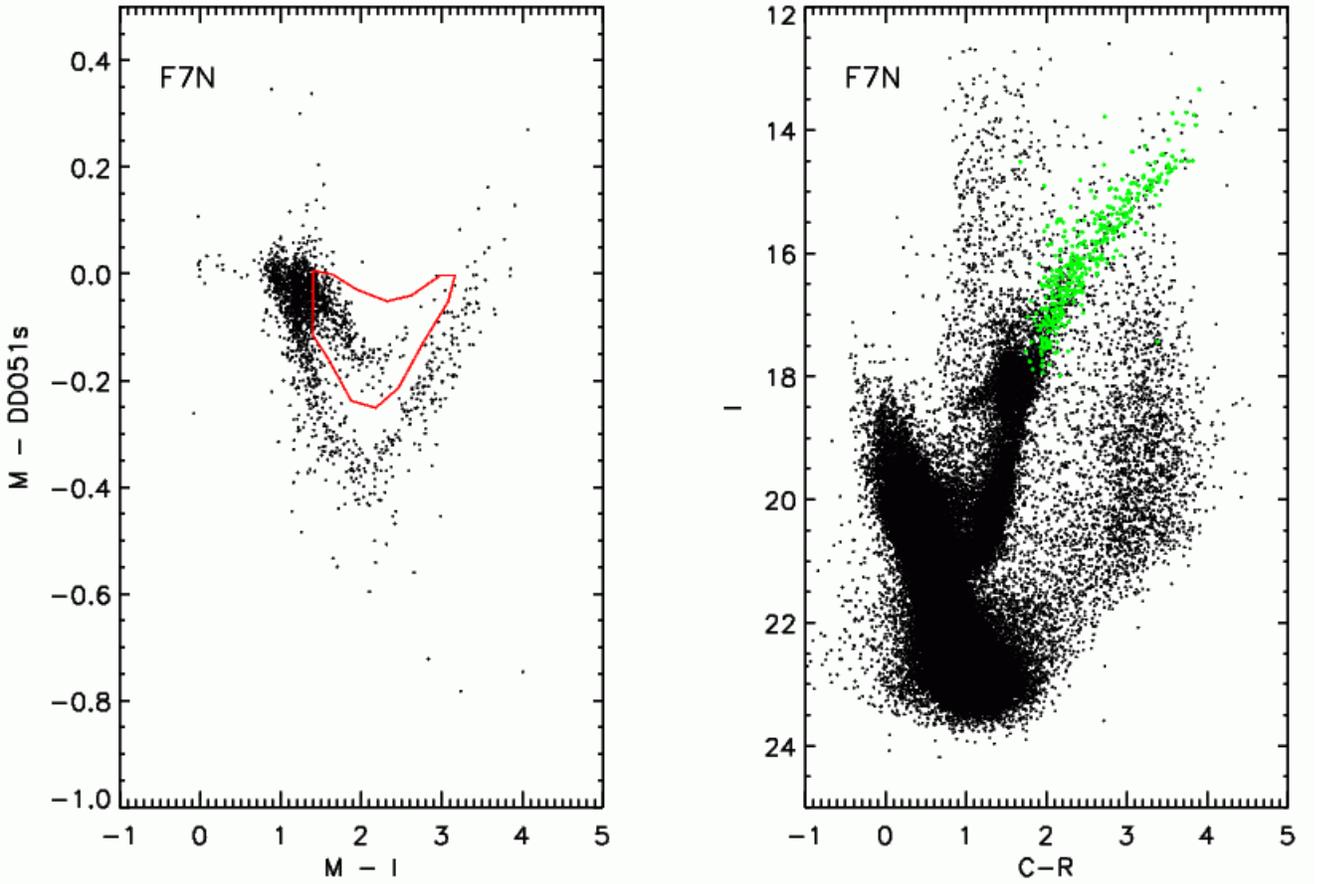}
\caption{The left panel shows a color-color diagram, $M-DDO51s$ vs $M-I$ for stars brighter than $I=18.0$.  The stars in the enclosed region are identified as candidate giants.  On the right hand panel, these candidate giants are shown in green on the CMD in $I$ vs $C-R$.  All of the objects fall in the region of the CMD containing the giants, demonstrating that this is a good and efficient way to identify giant candidates. See \S~\ref{sec_ddo51giants} for qualifications and caveats for this method.
\label{fig_F7ND51torgb}
}
\end{figure}

Consider Fig~\ref{fig_F7ND51torgb}.  The left panel shows the
color-color diagram of $M - DDO51s$ versus $M-I$ for stars brighter
than $I = 18$ (LMC giants brighter than the red clump cannot be
fainter than this), and where the reported uncertainties for $M$, $I$,
and $DDO51s$ are all less than 0.03 mag. At blue colors, or $M-I <
1.0$, all stars merge to $M-DDO51s \sim 0.0$, by construction.  At
redder colors, especially going past $M-I \sim 1.2$, we see separation
into two branches. The lower branch is populated by the foreground dwarfs, and the
upper branch corresponds to giants. For very much redder colors,
$M-I > 3.0$ the two branches merge again, and dwarf vs. giant
separation fails with this method at very low temperatures.  Stars
enclosed in the indicated region bounded in red are designed to include giants, and
reject dwarfs.

The right hand panel of Fig.~\ref{fig_F7ND51torgb} shows the now familiar CMD with $I$ vs $C-R$.  
The points shown in green are the same stars that enclosed in the `giants' area on the left hand panel. They lie on the giant branch location for the LMC giants, thereby demonstrating the efficacy of the method.  There are a few stars that follow the shape of the LMC giant branch, but lie above the 
visible concentration of RGB stars: these may be AGB stars, or perhaps they indicate the presence of complex structure in the RGB, possibly arising from the significant range of ages and metallicities  
indicated by the complex mixture evident from the MSTO region.  
A third possibility is that some of these
stars are from the Galaxy halo: we defer discussion of this to \S~\ref{sec_cmd-cts}.

It is notable that not all stars along the giant branch locus are marked in green.  Some of the unmarked ones are doubtless because they are really foreground dwarfs that happen to lie along that locus.
Notice that there are several unmarked stars near $C-R \sim 2.0 \pm 0.2$ and $I \sim 17.3$, where the density of points indicates that many of these must be LMC giants. Thus in this example we may have erred on the side of not including bona-fide giants.  Had we widened the enclosed area on the left hand color-color diagram, we might have included more giants, but at some point we would also begin 
to include non-giants.   The point is that this method can be used to identify giant candidates, which must later be followed up spectroscopically for confirmation. We should be circumspect about using this method to count giants,  because how we set our color-color limits, and the accuracy of photometry will govern the completeness as well as pollution of our giant sample.  As a pre-selection of objects for follow up spectroscopy, this method is excellent, but one should be wary of  making a stellar census from giants selected in this way.  

\begin{figure}[htbp]
\includegraphics[scale=0.6, angle=90]{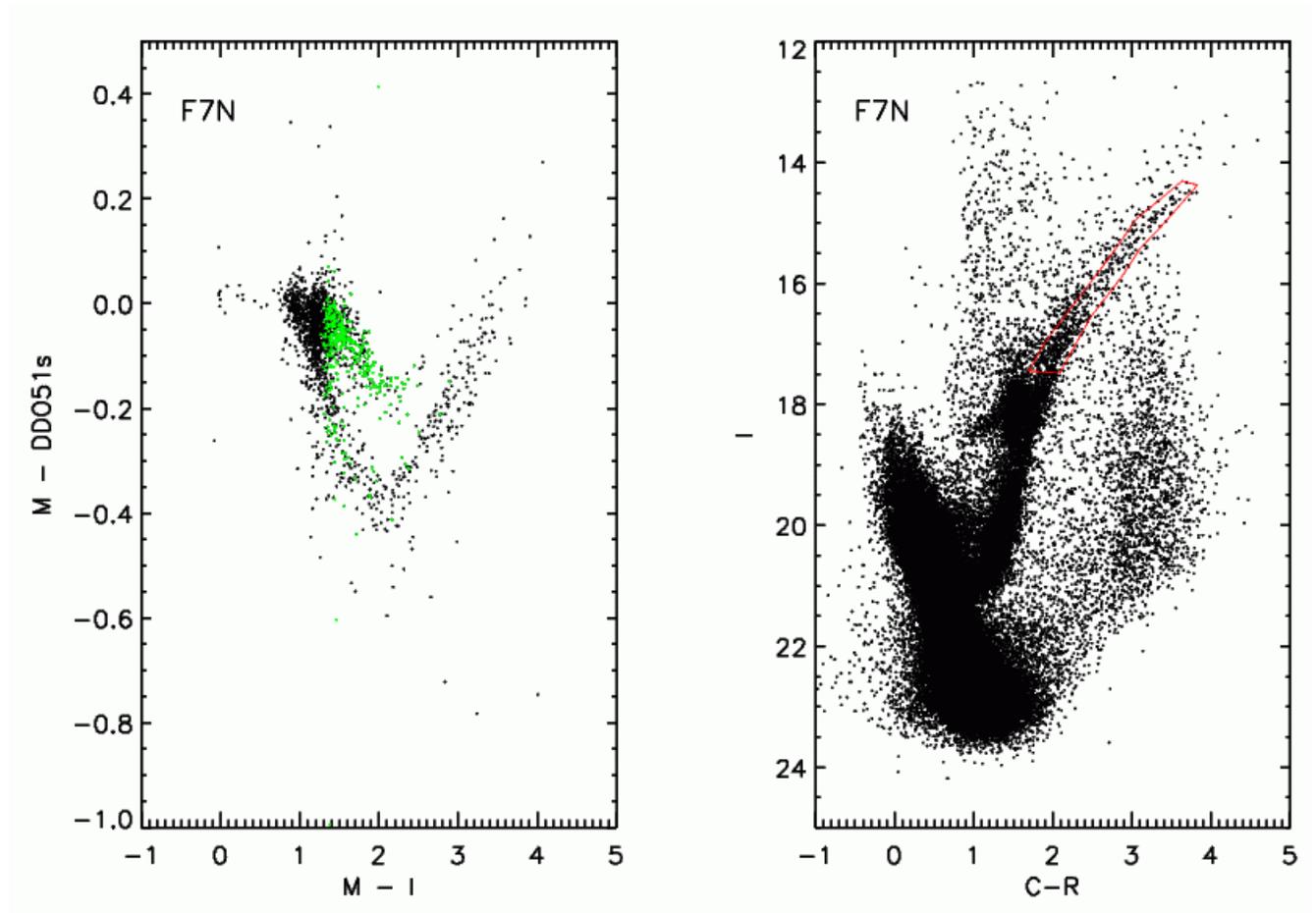}
\caption{Similar to Fig.~\ref{fig_F7ND51torgb}, but here the stars in the right hand panel enclosed in the 
RGB region  are mapped onto the color-color diagram in the left panel and shown as green points.  
While the majority of the points 
fall on the `giant branch' of the color-color diagram, a significant number though are clearly dwarfs, 
since the CMD region also has stars from the Galaxy foreground that are dwarfs.
 See \S~\ref{sec_ddo51giants} for a fuller discussion.
\label{fig_F7NrgbtoD51}
}
\end{figure}

Fig.~\ref{fig_F7NrgbtoD51} shows the reverse case, where a region of giants 
is chosen (very conservatively) on the CMD (shown in red on the right panel), and traced to the color-color diagram (left panel) where they are marked as green points (again, all points shown have 
reported photomteric errors less than 0.03 mag in $I$, $M$ and $DDO51s$). 
While the majority of the points 
fall on the `giant branch' of the color-color diagram, a significant number though are clearly dwarfs, 
since the CMD region also has stars from the Galaxy foreground that are dwarfs.  In fields where
the contribution from the LMC gets sparser, the marked region on the CMD will pick up more foreground
dwarfs than LMC giants.  Here also, one must independently assess the contribution of dwarfs
in the RGB region from control fields at similar Galactic latitudes.


\section{Analysis of Star Counts}
\label{sec_cmd-cts}

\begin{figure}[htbp]
\includegraphics[scale=0.7]{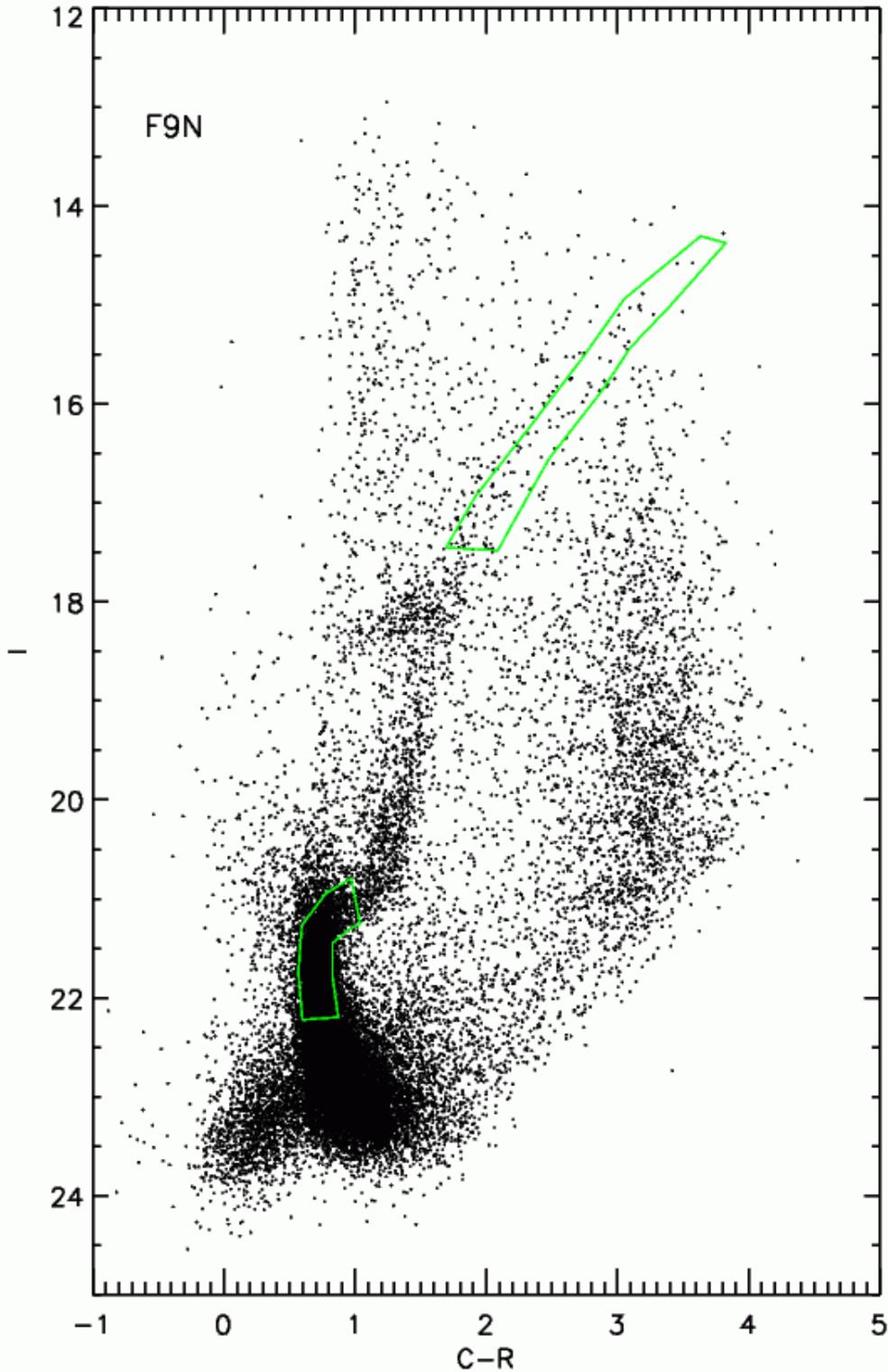}
\caption{The CMD of F9N is shown, on which two regions are marked. The lower one encloses stars 
with colors and brightness corresponding to the LMC main-sequence and old turn-off.  The upper region encloses the section of the CMD that contains RGB stars in the LMC brighter than the red clump:  this region was actually traced from the CMD of  F7N, where the RGB is better delineated.
\label{fig_regions} 
}
\end{figure}

Fig~\ref{fig_regions} shows the CMD of the field F9N again, but with
two regions marked.  The lower region encloses part of the lower main
sequence and the turn-off of the oldest stars. Its lowest (faintest)
extremities are chosen to be such that even in the worst seeing images
in any of the fields, the object detection is complete, and brighter
than the cloud of potentially unresolved galaxies mentioned above.
The higher region encloses the RGB stars brighter than the red clump
(its definition really comes from the CMD of F7N, where the RGB is
very clear).  These regions have been defined so that the numbers of
stars that lie within their boundaries can be counted and compared
across all the fields.

The control field C18 (see Table~\ref{tab_targfields}) is nominally a
perfect reference for estimating contamination for all of these fields
from foreground stars from the Galaxy, as well as for background
objects. However, this particular control field is itself
contaminated: we were surprised to find that it contains stars that
are an extension of the globular cluster NGC~1851, even though the
field lies several tidal radii from the cluster. This in itself was a
exciting discovery, and is reported elsewhere \citep{olz09} 
including a follow up investigation.  If we discard C18 as a result of
the above anomaly, we can use F121 and F122 as control fields {\it a
posteriori}, since they show no presence of stars associated with the
LMC.   

\begin{figure}[htbp]
\includegraphics[scale=0.7]{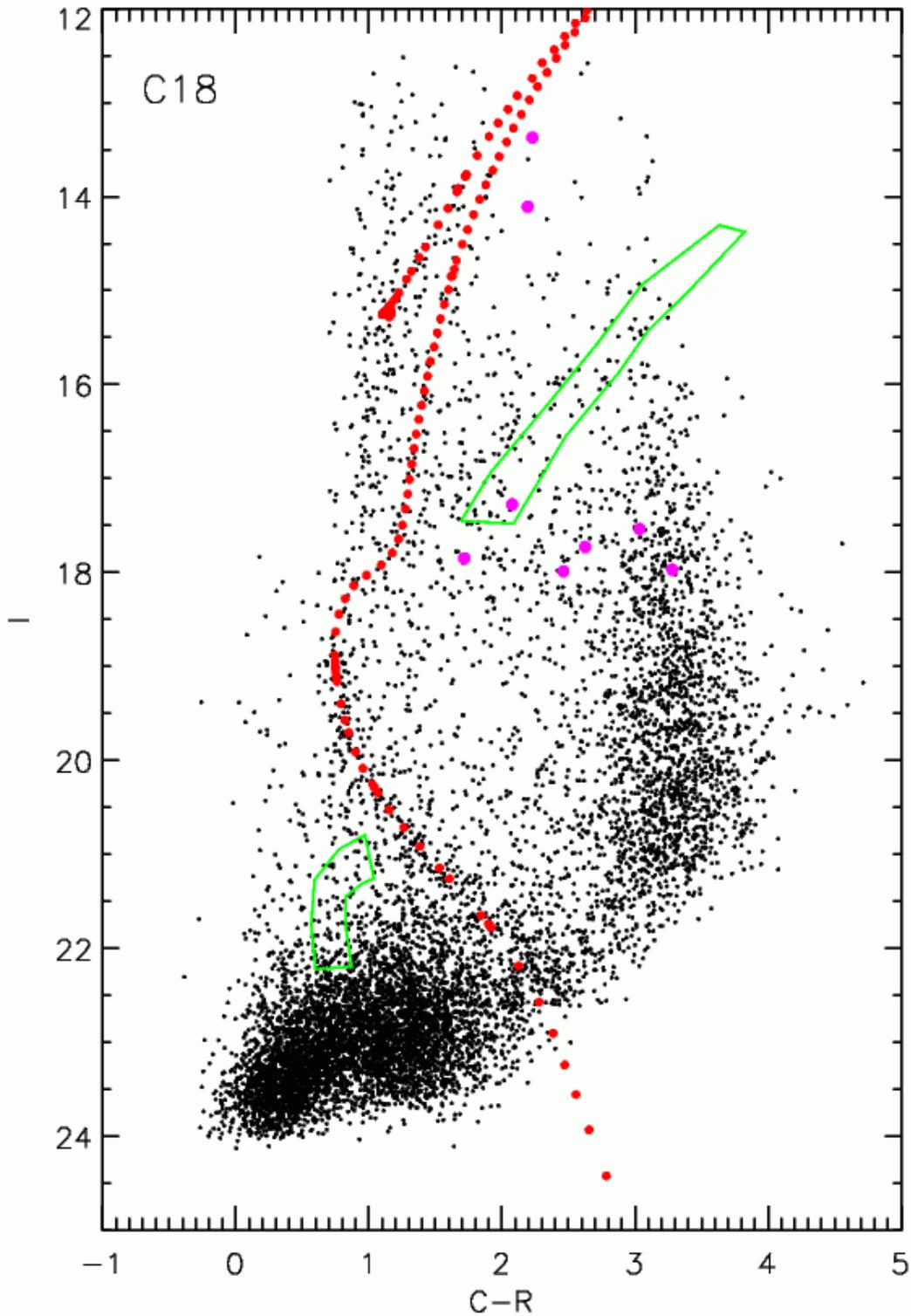}
\caption{\small The CMD of the field C18 is shown. This field is contaminated by the extended structure of 
the globular cluster NGC~1851: a locus of stars corresponding to the main-sequence of this cluster is visible. An isochrone suitable for NGC~1851 is over-plotted in red. The regions marked in green 
correspond to the LMC main-sequence and LMC RGB stars, as for Fig.~\ref{fig_regions}.
Note how both these regions should are clear of the isochrone, and hence not expected to be contaminated by stars belonging to NGC~1851.  The stars picked up as potential RGB candidates 
are shown by the purple filled circles.
}
\label{fig_C18cmd}
\end{figure}

It turns out, fortuitously,  that while stars associated with NGC~1851 are clearly present in the 
CMD of C18, they do {\it not} visibly pollute the two regions defined here.
 Fig.~\ref{fig_C18cmd} shows the $I$ {\it vs.} $C-R$ CMD 
for the field C18.  The feature corresponding to the main-sequence of 
NGC~1851 is clearly visible. An isochrone ($Z = 0.001$ and $\log t = 10.1$)
from \cite{Mar08} is over-plotted, using $E(B-V) = 0.02$ and $m-M = 15.2$, which 
matches the visible main-sequence.   Also over-plotted are the MS and RGB regions corresponding 
to the LMC CMD: it is clear that both these regions should be free from 
stars associated with NGC~1851, and are therefore useful for estimating the 
residual star density in these CMD regions.
We thus proceed  with caution to see if star counts from C18 can also 
help with background estimation.

The counts of stars in the two regions defined above for the various
fields being considered are presented in Table~\ref{tab_starcounts}.
In addition, the number of giants identified by the color-color diagram of 
$M - D51s$ vs $M - I$ described in \S~\ref{sec_ddo51giants} are listed in 
column (4). 
The three farthest fields (F121, F122 and C18) show an average of 73 stars in the MS
region, with scatter  within Poisson statistics.  Similarly the
number of stars in the RGB region averages to 50 for these 3 control
regions, again within reasonable Poisson statistical bounds.  It is
immediately clear that with the exception of F7N and F9N, there are no
significant excesses in the counts of RGB stars when compared to the
control value, confirming our visual examination based inference that
RGB stars run out as good tracers at distances larger than 8 or 9 degrees.
In comparison the counts of MS stars (obtained by subtracting the
average counts of stars in the MS locus on the CMDs for the three control
fields) in our pre-defined region can be traced out as far as 16
degrees (F123), with statistical significance, again corroborating our
visual examination based inference.

\begin{figure}[htbp]
\includegraphics[scale=0.8]{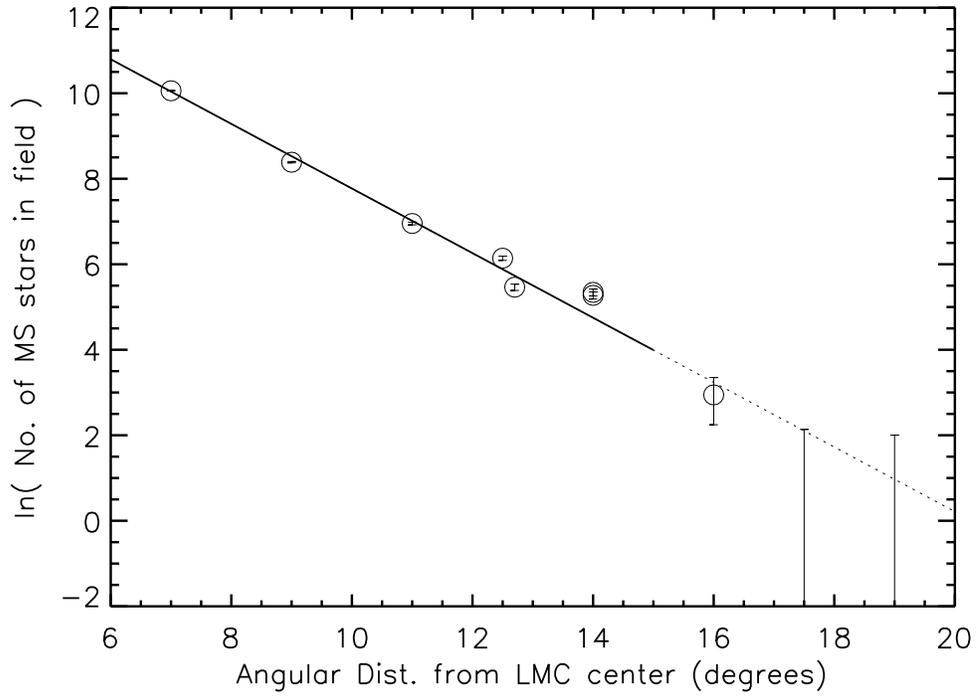}
\caption{The run of the log of surface density of MS and old TO stars selected from a pre-defined region of the CMD (see \S~\ref{sec_cmd-cts} for details) with distance as projected on the sky along a direction due North from the LMC center.  The line shown is a weighted fit to the inner 8 points, which implies an exponential decrease in the surface density of these stars with distance, with a scale length of  1.32 degrees {\it on the sky}.  \label{fig_expfit} 
}
\end{figure}

Fig~\ref{fig_expfit} shows the log of the surface density of stars in
the MS region from Table~\ref{tab_starcounts} (after subtracting the
average background value from the 3 control fields, following the discussion above)
against the projected distance from the LMC center.  The excellent fit
to an exponential decline in surface density over the entire range
where MS stars associated with the LMC can be detected formally favors
a pure disk, with a scale length of 1.32 degrees on the sky.
In \S~\ref{sec_chardisk}, we show that de-projecting onto the plane of 
the LMC disk yields a disk scale length of 1.15 kpc.  This value is remarkably close 
to that derived for the inner disk from counts of giants by \citet{vdm01}, who 
obtained a scale length   ``$ r_{d} ~\approx~ 1.3-1.5$ kpc.''  Other determinations
of the interior disk scale length,  $1.4-1.5$ kpc by \citet{bot88},
$1.42$ kpc by \citet{wein01}, and $1.47$ kpc by \citet{alv04}  are all mutually 
consistent.

\begin{figure}[htbp]
\includegraphics[scale=0.8]{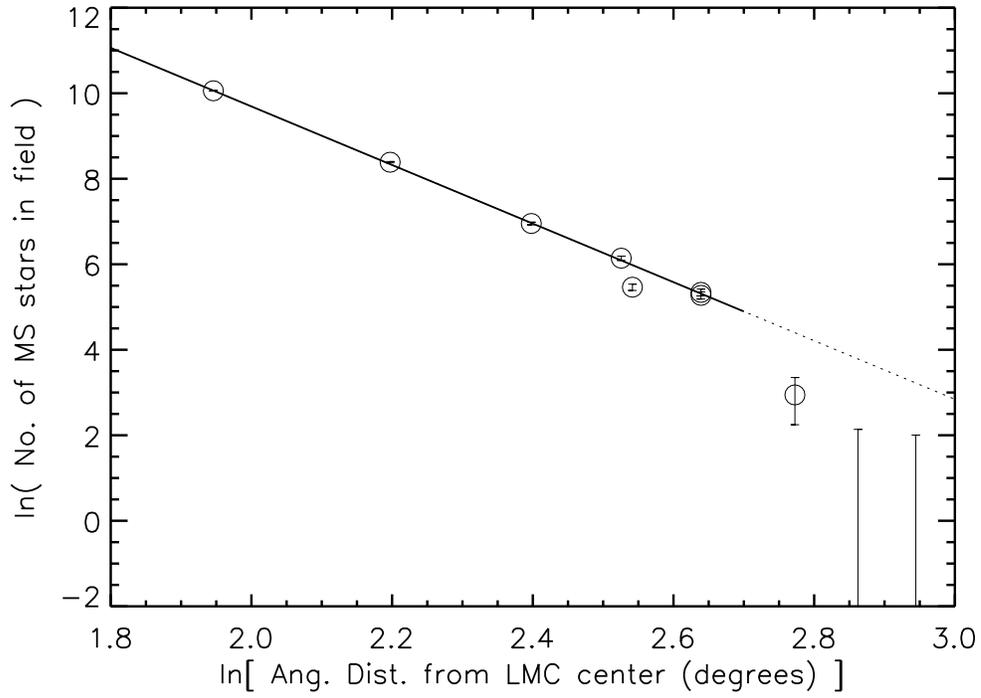}
\caption{  same as Fig~\ref{fig_expfit} , but with the abscissa showing the log of the distance from the LMC center.   A linear fit is possible only to the inner 7 points, and the implied power law  for surface density that results from such a weighted fit is  $\Sigma \propto R^{-6.85}$.  The unweighted fits is even steeper, and implies $\Sigma \propto R^{-7.03}$ \label{fig_powerfit}
}
\end{figure}

 In comparison, Fig~\ref{fig_powerfit} which a log-log plot of surface
 density vs. distance from the LMC center, is unable to fit the full
 range simultaneously, and even where the decline is most gentle,
 implies a power law $\Sigma \propto R^{-6.85}$.  In our own Galaxy,
 the halo has a much shallower radial dependence: \citet{sah85} showed
 that out to Galactocentric distance of 25 kpc, the density of RR
 Lyrae stars is consistent with $\propto R^{-3}$.  This analysis utilized 
 additional  data from  \citet{kin65},  and also showed that the `halo' is an 
 oblate spheroid near the Galaxy center, and becomes more spherical as 
 one goes out from the center.  
 Similarly \citet{zin85} obtained a density distribution for
 globular clusters that is $\propto R^{-3.5}$.  Since 
 a spatial power law that is $\propto R^{-n}$ implies a surface density 
 that is $\propto R^{1-n}$, these examples lead us to expect halo surface 
 density gradients that have power law indices near -2 or -2.5.  The index 
 implied above for the LMC extension is $-6.85$ which is very steep, and 
 too far a departure from our expectation to be a convincing model
 for a spheroidal halo.  In addition, given  the remarkable agreement of the exponential 
 disk scale length from our data with that from the prior value for  the inner disk, 
 in further discussion we exclude the possibility that we are seeing a spheroidal halo.
 
 The average number of stars in the RGB region (column 4 of Table~\ref{tab_starcounts})
 for the control fields F121, F122 and C18 is 50.  The corresponding star counts in 
 F11N through F14N and F123 are consistent with this value within $1-\sigma$ Poisson errors.
 The formal average is one RGB star per field, or equivalently 3 per square degree.
  Giants selected using the $D51s$ photometry 
 (column~5 in Table~\ref{tab_starcounts}) also follow the same trend.  
 
 We have verified that all of the stars in F121 and F122 that lie inside the nominal RGB box
 in the $C-R$ vs $I$ diagram, fall on the main sequence region of their respective $M - D51s$
 vs $M-I$ diagrams. Thus they cannot be RGB stars.  This also implies that 
 the small positive number of putative giants found by the $D51s$ method
 are either specious, or not related to the LMC.
 The counts in Cols.~4 and 5 of Table~\ref{tab_starcounts} for all fields except 
 F7N, F9N and C18 are similar enough to those in F121 and F122 
 within Poisson probabilities, that we can surmise that these fields are also free of LMC giants.
 Also note that when the `background' values are subtracted from the counts in Cols.~4 and 5
 (50 stars and 6 stars respectively), the number of implied LMC giants are quite similar in both 
 columns. Since each of these methods is afflicted in different ways (as discussed in 
 \S~\ref{sec_ddo51giants}) we find this rough agreement remarkable.   
 
 In the fields F7N and F9N, the ratio of  selected MS stars  to RGB selected giants (after 
 applying `background' corrections described above)  is greater than 50 (formally 54 in 
 F7N and 74 in F9N).   We can thus assert that every {\it bona fide} giant  in the LMC 
 must be accompanied by at least 50 dwarfs in our main sequence counting region.

\section{Discussion}
\label{sec_Discussion} 

\subsection{Further Characterization of the LMC Disk}
 \label{sec_chardisk}
 
 Having established that we are tracing the continuation of the disk characterized by \citet{vdm01}, 
 it is important that we set our observations in that context, and on that system,  especially to see where
 our fields lie on the plane of the LMC disk. 
 
 Whereas we used an LMC center with J2000 coordinates $\alpha ~ = ~ 5:23:34.0$ and $\delta ~ =  ~ -69:45:00$ 
 when designing our survey and calculating field positions, \citet{vdm01}  uses a projection origin $\alpha ~=~ 5:29$, 
 and $\delta ~=~ -69.5$.  This results in a small change to the angular distances of each of our fields 
 from the LMC center, and are tabulated in column (2) of Table~\ref{tab_vdmtable}. 
 The line of sight along a given field intersects the LMC disk at a distance that is 
 different from the distance to the LMC center, since the disk is tilted with the plane of the sky.
 These distances are calculated using the van der Marel disk geometry, and a fiducial distance to the LMC center of 
 50 kpc  (which corresponds to a distance modulus of 18.50) and listed in column~(3) of 
 Table~\ref{tab_vdmtable}. The corresponding distance modulus on the same basis is listed in column~(4).
 If the distance to the LMC center is denoted by $D_{0}$,   the line of sight distance to the disk by $D$, and if 
 the angular distance of this line of sight from the LMC center is $\theta$, then the distance $\rho$ along the plane of the disk 
 from the LMC center to the line of sight vector is given by:
 
 \begin{equation}
 \rho^2 ~=~ D^2 + D_{0}^2 - 2 D D_{0} \cos(\theta)
 \end{equation}
 
 Values of $\rho$ for each of the fields used here in the investigation of LMC structure 
are listed in column~(5)  of Table~\ref{tab_vdmtable}. 

\begin{figure}[htbp]
\includegraphics[scale=0.8]{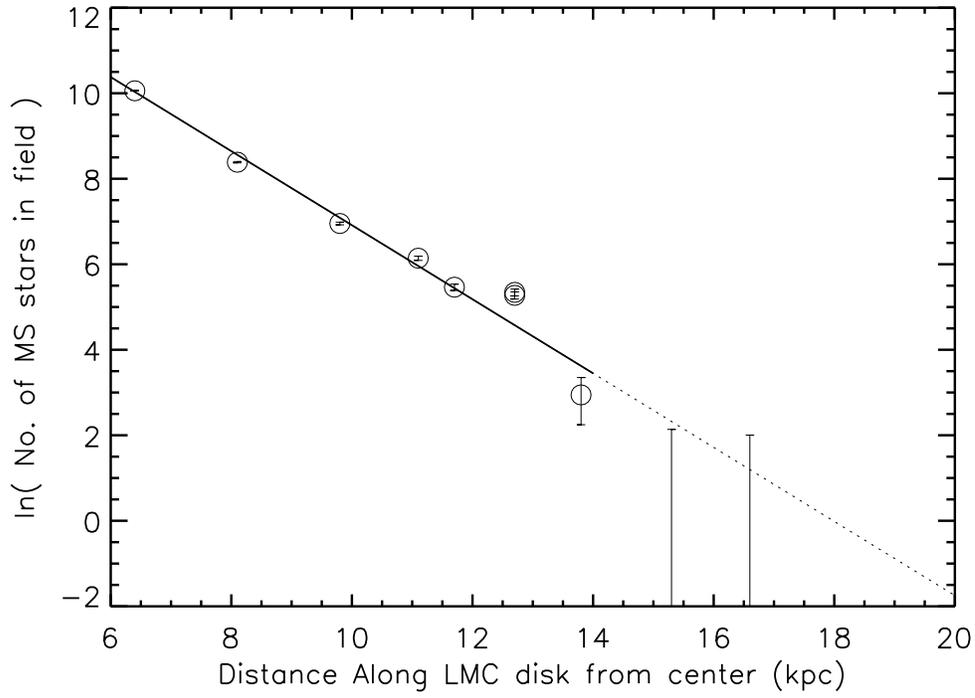}
\caption{Same as Fig.~\ref{fig_expfit}, but with the abscissa values showing distance along the plane of the 
LMC disk using the disk geometry from \citet{vdm01}, and a fiducial distance to the LMC center of 50 kpc.  A weighted linear fit to the inner 8 points yields a disk scale length of 1.15 kpc, and is shown by the line.
 An unweighted fit gives 1.20 kpc.
 \label{fig_diskfit} 
}
\end{figure}

Fig.~\ref{fig_diskfit} shows the run of surface density against true LMC-centric distance along  van der Marel's 
LMC disk.  We have ignored any effects that disk warping and flaring would cause.  Detection of 
such features  in our data would require more complete spatial coverage.

\subsection{Can there be an undetected LMC halo?}
\label{sec_possiblehalo}

We have shown that the LMC disk is traceable to $16^{\circ}$, which is more than 10 scale lengths.
We must ascertain at what level our results rule out the existence of an LMC halo.
We can first ask what one would see if a
similar experiment were directed at the Milky Way Galaxy, from a
vantage point from which the plane of the Galaxy is seen face-on.
We recognize that the literature provides a vast array of models 
for the Milky Way disk, thick disk, halo and bulge.  Each is based on different 
tracers for these components, which in turn are normalized to all stars 
using IMFs and spatial dependences.  Each  approach has strong 
and weak points.  A proper discussion of the state 
of knowledge is beyond the scope of this paper, but the interested 
reader can see, for example: \citet{bah84}, \citet{sanfou87}, \citet{san87}, 
\cite{mor00} \& \cite{rob03}.
Instead we make a few 
heuristic assertions that we deem reasonable, and proceed towards the goal 
of getting an approximate picture of the radial run of star count surface densities 
would look like if the Galaxy were viewed face on.

Accordingly we adopt from from 
 \citet{Dri01} the Milky Way radial disk scale length $r_{GAL} = 0.28D_{\sun}$,  
 where $D_{\sun}$ is the Galactocentric distance of the Sun. We assert that it is adequate 
 for our purpose to assume that  this applies to both the thin and thick disks.
 For the halo we adopt a simple spherical distribution with a power law density 
 dependence of $R^{-3.5}$, which is borne out by the density distribution of globular 
 clusters \citep{zin85}  and consistent with the distribution of RR Lyrae stars
  \citep{sah85}.  This corresponds to a surface density distribution that is $\propto R^{-2.5}$.
 We query the Besancon model \citep{rob03} for counts of stars with $3.5 \leq M_{V} \leq 6.0$
 towards the north and south Galactic poles, segregated by thin disk, thick disk, halo and bulge components, from which we can derive the approximate column densities of these stars. The luminosity cut is representative of the main sequence dwarfs that we have used to trace surface density in the 
 LMC exterior.  The counting of stars is done along a cylindrical column, and not as seen 
 in through an opening with a fixed solid angle. In this way, we derive a total column density 
 of 3.56 , 1.12, and 0.03 stars $pc^{-2}$ for the thin disk, thick disk, and halo components respectively, 
 at $D_{\sun}$, and zero contribution from the bulge.    These component column densities  are 
 identical to  the surface densities from these components as seen by a hypothetical distant observer looking face-on at the Galaxy.  We use these surface densities at the solar location in the Galaxy, we can 
invoke  our simplistic disk-halo model, to obtain expressions for the run of disk and halo surface 
densities ($\Sigma_{disk}$ and $\Sigma_{halo}$ respectively):
 \begin{equation}
 \Sigma_{disk}  = 137.3 ~e^{-D_{\sun}/0.28}  
\label{eq_sigdiskgal}
\end{equation}
and
\begin{equation}
\Sigma_{halo} = 0.030~ D^{-2.5}_{\sun}
\label{eq_sighalogal}
\end{equation}

\begin{figure}[htbp]
\includegraphics[scale=0.8]{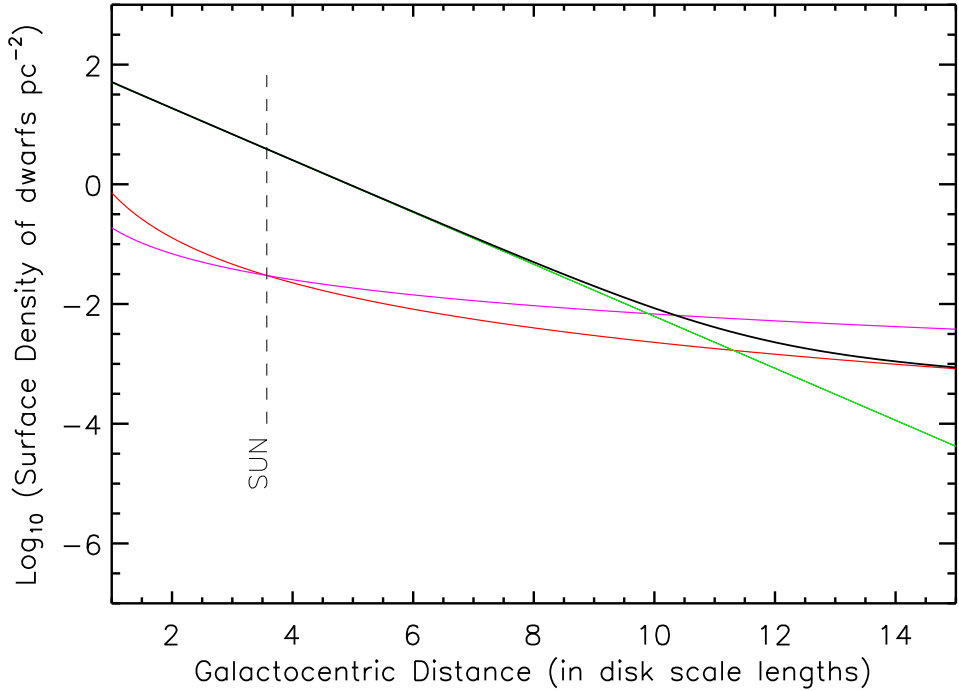}
\caption{The run of surface star density for dwarfs in the luminosity range $3.5 \leq M_{V} \leq 6.0$, 
that would appear to an observer viewing the Milky Way face-on from a large distance. The green line shows the contribution from the disk with scale length 0.28 times the Galactocentric distance of the Sun. Both thin and thick disk components are included.  The red line shows the contribution from an $R^{-3.5}$ halo, whereas the magenta line shows the contribution if an 
$R^{-2.44}$ halo is assumed instead: they are normalized at the position of the Sun.  The black line shows the combined contribution of stellar surface density of stars in the disk plus an  $R^{-3.5}$ halo.  For the details of the model, see \S~\ref{sec_Discussion}. Irrespective of model differences,  the overall conclusion is that in the Milky Way, the halo does not overtake the disk  till about 10 disk scale lengths, or about 3 times the Galactocentric distance of the Sun.
\label{fig_MW-faceon}
}
\end{figure}

Fig.~\ref{fig_MW-faceon} plots the relative contribution from disk and halo components as a function of distance along the disk.  The disk continues to dominate till about 10 disk scale lengths. Thus an 
observer looking face on at the Milky Way would see a disk like exponential fall off in star counts all the 
way to 10 scale lengths.  While there is no reason at all to believe that the proportion of disk and halo 
should scale from one galaxy to another, nevertheless this indicates that our non-detection of 
a halo in the LMC does not mean that it does not have one, since we know that the Galaxy does 
have a halo, but that it does not nominally become prominent until past 10 scale lengths (roughly 25 kpc).  Of course other issues such as disk warping, and termination (tidal or otherwise) of either disk, or halo or both in the outer parts can complicate this simplistic scenario.

We can also compare the actual surface density values expected in  a face-on viewing of the Galaxy
with what we see in the LMC.   In F11, we count about 1050 dwarfs belonging to the LMC
(Table~\ref{tab_starcounts}).  The footprint of the MOSAIC2 field at the LMC distance is $\sim 0.25 ~ {\rm kpc}^{2}$, thus implying a surface density of LMC dwarfs (picked from our CMD region) of 
$\sim 0.004 ~ {\rm pc}^{-2}$.   F11N is at about 8 LMC disk scale lengths.  The equivalent location 
of 8 disk scale lengths in our Galaxy is at $ \sim 2.3 ~ D_{\sun}$, where the density of 
dwarfs with luminosity $3.5 \leq M_{V} \leq 6.0$ (i.e. commensurate with the dwarfs identified from 
our CMD region in the LMC) from the Galaxy disk is $\sim 0.04 ~ {\rm pc}^{-2}$, with the halo 
contribution an order of magnitude or so lower.   

A natural question to ask is what limits our data can place on the detection of an LMC halo 
yet farther out using the detection of RGB stars yet farther out,  \citep{nid07,maj99, maj09}.
Our outermost detection of dwarfs is at 16 degrees from the LMC in F123, 
with a dwarf surface density of $\sim 20$  per MOSAIC field, which translates to 
$8 \times 10^{-5} ~{\rm pc}^{-2}$.   Let us, for the sake of argument,  take the extreme position 
that at this location it is the halo, rather than the disk which is the main contributor, and further that
the halo continues with a {\it flat} density out to the LMC distances reported by the above works.
Their fields, which extend out to 23 degrees from the LMC center, would have an upper limit for the surface count density of $8 \times 10^{-5} ~{\rm pc}^{-2}$.
We have established in \S~\ref{sec_cmd-cts}, that there must be at least 50 dwarfs for each RGB star 
seen, which therefore places an upper limit of $1.6 \times 10^{-6}  ~{\rm pc}^{-2}$ or 
$1.6  ~ {\rm kpc}^{-2}$ for the column density 
of RGB stars, using our already extreme model for a putative LMC halo.  This upper limit corresponds to 
one RGB star belonging to the LMC per square degree.  An $R^{-3.5}$ halo instead of a flat continuation would imply one RGB star every 2.5 square degrees  at a field 23 degrees from the LMC center.

\citet{mun06} discovered a set of 15 giant stars roughly centered on Carina that have a velocity
consistent with a smoothly changing LMC velocity and a velocity dispersion of 9.8 km/s.
The density of these stars is  $ \sim 1~{\rm kpc}^{-2} $,  or $\sim 25$ per MOSAIC field, which is the same density as in our 16 degree field.
The distribution on the sky of these stars, discovered from a 4x4 degree imaging survey is not
smooth (see Fig 17 in Munoz et al), and we speculate that it is consistent with stripping
from the LMC disk itself.

 An $R^{-3.5}$ halo that overtakes the disk at  16 degrees from the LMC center predicts 16 dwarfs 
 in our F122 field, and 13 in the F121 field.  We have found a total of 143 objects 
 (from Table~\ref{tab_starcounts}) in the MS region of the 
 CMD (with no back/fore-ground correction) in these two fields taken  together. For 29 of these 
 to be LMC stars, the fore-background counts in the MS region would have to be about 55 per field.
 We cannot altogether rule this out, but the fact that we see 78 total objects in the MS region 
for C18, which is 34 degrees away, does not favor such a possibility. Besides, the 20 dwarfs
seen in F123 delineate a faint MS without difficulty,  and so it is quite unlikely that 16 dwarfs 
in F122 would entirely escape being noticed by eye.

 \subsection{Summary}
 
 We have described our survey, and the data processing.  We have
demonstrated in this paper that the data are of adequate quality for
addressing the questions that have motivated this survey.

We have presented the results for a set of fields extending due North
from the LMC, where we are able to trace stars associated with the LMC
out past 10 disk scale lengths, which is unprecedented for any galaxy
to date. This corresponds to surface brightness levels of 34 mag
arcsec$^{-2}$.

The observed surface density of stars is consistent with a spatial distribution 
due to an exponential disk with scale length $\approx 1.15$ kpc, which agrees with 
previous determination of the scale of the LMC disk interior to 9 kpc.
We are unable to detect a contribution from a
spheroidal halo component. Rather, we note  that out along this direction due North from the 
LMC center,  the structure of the disk appears perfectly disk-like
and unperturbed out to $\sim 16$ kpc. There are no noticeable effects from tidal stripping.  Of course, this 
direction was chosen because, by running parallel to the Galactic plane, and by being in the direction
opposite to the SMC, it was expected to be the least affected by tides.   Even so, our result is noteworthy, 
since this line of fields goes
twice as far out from the LMC compared to the region where tidal streamers in HI and in carbon stars have been reported by \citet{ols07}.  It goes without saying that scrutiny of  results from future analyses in other fields and regions of our survey will hold great interest.

In future papers we will present both, similar analysis of data from other fields, as well as detailed 
population analyses of Hess diagrams in the various locations that we have sampled.

\acknowledgments

We thank the CTIO staff,especially the telescope 
operators and technical staff, for their help on the many observing runs. We also thank the NOAO archive group for making it easy to retrieve
our data from the Tucson archive, and for their help with getting our final processed images ready for
the public archive.  Saha thanks Frank Valdes for illuminating discussions about image processing
in general, and IRAF help in particular.
EO acknowledges NSF grant AST-0807498.
BB gratefully acknowledges fellowship support from the Indiana Space Grant Consortium.
DM is supported by grants from FONDAP CFA 15010003, BASAL CATA PFB-06, and MIDEPLAN MWM P07-021-F.
AR thanks the NOAO Goldberg Fellowship Program for its support

\clearpage



\clearpage

\clearpage


\clearpage


\clearpage

\clearpage

\clearpage

\clearpage

\clearpage

\clearpage

\clearpage

\clearpage

\clearpage

\clearpage

\clearpage

\clearpage

\clearpage

\clearpage

\begin{deluxetable}{lcccc}
\tabletypesize{\scriptsize}
\tablecaption{Designations and Central Coordinates of Survey Fields \label{tab_targfields}}
\tablewidth{0pt}
\tablehead{
\colhead{Field Name}  & \colhead{Gal. Long.} & \colhead{Gal. Lat.} & \colhead{RA (J2000)} & \colhead{DEC (J2000)} \\
\colhead{} & \colhead{$\ell^\circ$} & \colhead{$b^\circ$} & \colhead{(hh:mm:ss)} & \colhead{ (~$^{\circ}$ :~  '  :~  ") }
}
\startdata
  C1     &   330  &    -15    &     17:34:00   &      -62:36:00     \\
  C2     &   330  &    -30    &     19:45:40   &      -65:36:00      \\
  C3     &   330  &    -45    &     22:00:40   &      -61.42:00      \\
  C4     &   330  &    -60    &     23:20:26   &      -52:42:00      \\
  C5     &     0  &    -75    &     23:50:40   &      -34:00:00       \\
  F506   &   310  &    -45    &     23:55:00   &      -71:20:00       \\
  F504   &   313  &    -45    &     23:25:00   &      -70.24:00       \\
  F503   &   316  &    -45    &     23:03:52   &      -69.18:00        \\
  F502   &   319  &    -45    &     22:52:43   &      -68:05:45        \\
  F501   &   322  &    -45    &     22:35:28   &      -67:06:42     \\
  F307   &   301  &    -55    &      1:05:58   &      -62:31:59       \\
  F305   &   301  &    -57    &      0:57:02   &      -60:10:02       \\
  F304   &   301  &    -66    &      0:56:17   &      -51:00:42       \\
  F306   &   300  &    -70    &      0:59:52   &      -47:20:19       \\
  F308   &   299  &    -57    &      1:13:34   &      -60:33:50       \\
  F309   &   295  &    -57    &      1:21:27   &      -59:34:45      \\
  F301   &   308  &    -58    &      0:30:51   &      -58:37:33      \\
  F508   &   310  &    -50    &      0:04:31   &      -66:22:33       \\
  F507   &   307  &    -45    &      0:04:02   &      -70:59:48      \\
  F521   &   309  &    -47    &      0:05:00   &      -69:35:00       \\
  F4C1   &   290  &    -20    &      8:48:49   &      -76:14:31     \\
  F4C4   &   294  &     -8    &     11:10:52   &      -68:56:34     \\
  F4C6   &   285  &    -15    &      8:58:34   &      -69:22:14     \\
  F411   &   300  &    -20    &     11:35:08   &      -82:36:43     \\
  F412   &   303  &    -15    &     12:47:08   &      -77:59:30     \\
  F414   &   308  &     -8    &     13:45:33   &      -70:31:32     \\
  F415   &   310  &     -5    &     14:06:40   &      -66:45:55     \\
  F404   &   295  &    -25    &      7:49:36   &      -82:46:02     \\
  F405   &   295  &    -20    &      9:43:47   &      -80:01:53     \\
  F111   &   263  &    -35    &      5:20:25   &      -55:20:15     \\
  F113   &   265  &    -31    &      5:48:49   &      -57:02:20     \\
  F7N    &   272  &    -34    &      5:23:34   &      -62:45:00     \\
  F9N    &   269  &    -34    &      5:23:34   &      -60:45:00     \\
  F11N   &   267  &    -34    &      5:23:34   &      -58:45:00     \\
  F12p5N &   265  &    -34    &      5:23:34   &      -57:15:00     \\
  F14N   &   263  &    -34    &      5:23:34   &      -55:20:00     \\
  C20    &   245  &    -25    &      6:04:38   &      -38:25:35     \\
  C18    &   245  &    -35    &      5:15:20   &      -40:43:05     \\
  C14*   &   245  &    -55    &      3:28:35   &      -40:23:56     \\
  C12    &   245  &    -65    &      2:40:32   &      -38:03:50     \\
  C102   &   225  &    -75    &      1:58:00   &      -29:30:00     \\
  F605   &   283  &    -50    &      2:45:57   &      -62:39:45     \\
  F603   &   290  &    -48    &      2:21:24   &      -66:54:24     \\    
  F601   &   297  &    -43    &      1:53:46   &      -73:36:10        \\
  F531   &   312  &    -45    &     23:32:08   &      -70:41:27      \\
  F532   &   310  &    -43    &     23:41:16   &      -73:08:22        \\
  F533*   &   310  &    -47    &     23:55:20   &      -69:28:35        \\
  F534   &   304  &    -46    &      0:43:21   &      -70:54:00        \\
  F121*  &   258  &    -34    &      5:21:50   &      -50:28:16      \\
  F122   &   260  &    -34    &      5:21:59   &      -52:05:33      \\
  F123   &   258  &    -34    &      5:24:43   &      -54:02:27        \\
  F141   &   269  &    -26    &      6:26:15   &      -59:39:12        \\
  F142   &   268  &    -25    &      6:36:55   &      -58:47:40      \\
  F143   &   266  &    -23    &      6:40:55   &      -56:50:34      \\
  F144*   &   265  &    -22    &      6:50:45   &      -55:14:38         \\
\enddata
\tablenotetext{a}{Fields marked with * have missing observations, but the partial available data are useful enough}
\end{deluxetable}




\clearpage

\begin{deluxetable}{ccccccc}
\tabletypesize{\scriptsize}
\tablecaption{Example Photometric Solution (for 2007 Oct 12)  \label{tab_ntcoeffs}}
\tablewidth{0pt}
\tablehead{
\colhead{Pass Band}  & \colhead{$~~~\alpha~~~$} & \colhead{$~~~\beta~~~$} & \colhead{$~~~\gamma~~~$} & \colhead{COLOR}  & \colhead{rms residual} & \colhead{No. of Obs.}  \\
}
\startdata
R & 7.573 & 0.080 & -0.014 & R-I   & .018 & 136 \\
R & 7.576 & 0.081 & -0.003 & C-R &  .014 & 67 \\
I   & 8.427 & 0.039 & -0.018 & R-I   &  .020  & 130 \\
C & 8.021 & 0.312 & -0.021 & C-R  &  0.019 & 68 \\
M & 6.901 & 0.142  & -0.015 & M-R & 0.020 & 66 \\    
\enddata
\end{deluxetable}

\clearpage

\begin{deluxetable}{cccccc}
\tablecaption{Chip to Chip response variations in MOSAIC2 on 2005 Oct 3 (UT) \label{tab_colresponse}}
\tablewidth{0pt}
\tablehead{
\colhead{Pass Band} & \colhead{CCD Chip No.}  & \colhead{${\cal A}^\prime$} & \colhead{${\cal B}^\prime$} & \colhead{rms residual} & No. of stars used  \\
}
\startdata
R & 1 & -4.276 & .076 & .011 & 6 \\ 
    & 2 & -4.297 & .045 & .012  & 6 \\
    & 3 & -4.320 & .039 & .012  & 6 \\
    & 4 & -4.289 & .078 & .012  & 6 \\
    & 5 & -4.292 & .019 & .010  & 6 \\
    & 6 & -4.297 & .016 & .011  & 6 \\
    & 7 & -4.298 & .019 & .013  & 6 \\
    & 8 & -4.295 & .012 & .013  & 6 \\
    & $<$all CCDs$>$ & -4.296 & .038 & .017 & 48 \\
  \\
 I & 1 & -4.997 & -.028 & .020 & 6 \\ 
    & 2 & -5.001 & -.011 & .017  & 6 \\
    & 3 & --5.043 & -.022 & .015  & 6 \\
    & 4 & -5.003 & -.029 & .018  & 6 \\
    & 5 & -5.008 & -.096 & .018  & 6 \\
    & 6 & -5.012 & -.069 & .020  & 6 \\
    & 7 & -5.015 & -.066 & .018  & 6 \\
    & 8 & -5.016 & -.063 & .016  & 6 \\
    & $<$all CCDs$>$& -5.012 & -.047 & .022 & 48 \\
  \\
  C & 1 & -5.168 & .044 & .011 & 6 \\ 
    & 2 & -5.204 & .040 & .008  & 6 \\
    & 3 & -5.158 & .034 & .011  & 6 \\
    & 4 & -5.196 & .047 & .009  & 6 \\
    & 5 & -5.132 & .028 & .020  & 6 \\
    & 6 & -5.125 & .025 & .013  & 6 \\
    & 7 & -5.218 & .043 & .006  & 6 \\
    & 8 & -5.200 & .038 & .010  & 6 \\
    & $<$all CCDs$>$ & -5.165 & .037 & .022 & 48 \\  
\enddata
\end{deluxetable}

\clearpage

\begin{deluxetable}{lcccc}
\tablecaption{Numbers of stars in selected CMD regions in the LMC extension fields \label{tab_starcounts}}
\tablewidth{0pt}
\tablehead{
\colhead{Field Name} & \colhead{Dist. from LMC center} & \colhead{No. in MS region} &   \colhead{No. in RGB region} & \colhead{No. of D51s}  \\
\colhead{} & \colhead{(degrees on sky)} & \colhead{} &  \colhead{}  & \colhead{selected giants}  \\
\colhead{(1)} & \colhead{(2)} & \colhead{(3)} & \colhead{(4)}  & \colhead{(5)} \\
}
\startdata
F7N & 7.0 &  23409 &  532 & 463 \\
F9N & 9.0 &    4462  & 109  & 54  \\
F11N & 11.0 & 1119 & 55 & 4  \\
F12p5N & 12.5 & 537 & 55 & 2 \\
F14N & 14.0 & 281 & 47 & 11 \\
F123 & 16.0 & 91 & 45 & 2 \\
F122 & 17.5 & 72 & 49 & 3 \\
F121 & 19.0 & 71 & 38 & 5 \\
C18 & 34.0 & 78 & 63 & 9 \\
F111 & 14.0 & 267 & 52 & 9 \\
F113 &  12.7 & 308 & 65 & 16 \\ 
\enddata
\end{deluxetable}

\clearpage

\begin{deluxetable}{lcccc}
\tablecaption{Position of Fields wrt van der Marel's LMC disk geometry
\label{tab_vdmtable}
}
\tablewidth{0pt}
\tablehead{
\colhead{Field ID} & \colhead{Angular Distance} & \colhead{Line of Sight} & \colhead{(m-M)} & \colhead{Distance along} \\
\colhead{} & \colhead{from vdM disk center} & \colhead{Distance} & \colhead{} & \colhead{vdM LMC Plane} \\
\colhead{} & \colhead{(degrees)} & \colhead{(kpc)} & \colhead{} & \colhead{(kpc)} \\
\colhead{(1)} & \colhead{(2)} & \colhead{(3)} & \colhead{(4)}  & \colhead{(5)} \\
}
\startdata
F7N     &     6.77   &    47.3 &   18.37  &   6.4 \\
F9N     &     8.77   &    46.6 &   18.34  &   8.1 \\
F11N    &    10.77   &    46.0 &   18.31  &   9.8 \\
F12p5N  &    12.26   &    45.6 &   18.30  &  11.1 \\
F14N    &    14.18   &    45.2 &   18.27  &  12.7 \\
F121    &    19.05   &    44.3 &   18.23  &  16.6 \\
F122    &    17.43   &    44.6 &   18.25  &  15.3 \\
F123    &    15.47   &    44.8 &   18.26  &  13.8 \\
C18     &    28.84   &    43.9 &   18.21  &  24.1 \\
F111    &    14.20   &    45.3 &   18.28  &  12.7 \\
F113    &    12.65   &    44.7 &   18.25  &  11.7 \\
\enddata
\end{deluxetable}

\clearpage





\begin{thebibliography}{}

\bibitem[Alves (2004)]{alv04} Alves, D. R. 2004, \apjl, 601, L151
\bibitem[Bahcall \& Soneira (1985)]{bah84} Bahcall, J. N. \& Soneira, R. M. 1984, \apjs, 55, 67
\bibitem[Bothun \& Thompson(1988)]{bot88} Bothun, G.~D., \& Thompson, I.~B.\ 1988, \aj,96, 877 
\bibitem[Canterna (1976)]{can76} Canterna, R. 1976, \aj, 81, 228
\bibitem[Clark \& McClure (1979)]{cla79} Clark, J. P. A. \& McClure, R. D. 1979, \pasp, 91, 507
\bibitem[Dolphin et al. (2001)]{dol01} Dolphin, A. E., Walker, A. R., Hodge, P. W., Mateo, M., Olszewski, E. W., Schommer, R. A., \& Suntzeff, N. B. 2001, \apj, 562, 303
\bibitem[Drimmel \& Spergel (2001)]{Dri01} Drimmel, R., \& Spergel, D. N. 2001, \apj, 556, 181
\bibitem[Feast (1968)]{fea68} Feast, M. W. 1968, \mnras, 140, 345
\bibitem[Freeman et al.(1983)]{fre83} Freeman, K.~C., Illingworth, G., \& Oemler, A., Jr. 1983, \apj, 272, 488 
\bibitem[Gallart et al. (2004)]{gal04} Gallart, C., Stetson, P. B., Hardy, E., Pont, F., \& Zinn, R. 2004, \apj, 614, L109
\bibitem[Gallart et al. (2008)]{Gal08} Gallart, C., Stetson, P.B., Meschin, I.P., Pont, F., \& Hardy, E. 2008, \apj, 682, L89
\bibitem[Geisler (1984)]{gei84} Geisler, D. 1984, \pasp, 96, 723
\bibitem[Geisler (1996)]{gei96} Geisler,  D. 1996,  \aj, 111, 480
\bibitem[Geisler (2005)]{gei05} Geisler, D.  2005, priv. comm.
\bibitem[Harris(2007)]{har07} Harris, J.\ 2007, \apj, 658, 345
\bibitem[Irwin (1991)]{irw91} Irwin, M. 1991, in The Magellanic Clouds: Proceedings of IAU Symposium 148, Kluwer Academic Publishers, eds. R. Haynes \& D. Milne, p.  453
\bibitem[Kallivayalil et al. (2006)]{kal06} Kallivayalil, N., van der Marel, R. P., Alcock, C., Axelrod, T., Cook, K. H., Drake, A. J., \& Geha, M. 2006, \apj, 638, 772
\bibitem[Kim et al. (2003)]{kim03} Kim, S., Staveley-Smith, L.,  Dopita, M. A., Sault, R. J., Freeman, K. C., Lee, Y., \& Chu, Y. 2003, \apjs, 148, 473
\bibitem[Kinman et al. (1965)]{kin65} Kinman, T.D., Wirtaanen, C. A., \& Janes, K. A. 1965, \apjs, 11, 223 
\bibitem[Kinman (1991)]{kin91} Kinman, T. D., Stryker, L. L., Hesser, J. E., Graham, J. A., Walker, A. R., Hazen, M. L., \& Nemec, J. M. 1991, \pasp, 103, 1279 
\bibitem[Kunkel et al. (1997a)]{kun97} Kunkel, W.~E., Demers, S., Irwin, M.~J., \& Albert, L.\ 1997, \apjl, 488, L129 
\bibitem[Kunkel et al. (1997b)]{kun97b} Kunkel, W.~E., Irwin, M.~J., \& Demers, S. 1997, \aaps, 122, 463
\bibitem[Kunkel et al. (2000)]{kun00} Kunkel, W.~E., Demers, S., \& Irwin, M.~J.\ 2000, \aj, 119, 2789 
\bibitem[Landolt (1983)]{lan83} Landolt, A. 1983, \aj, 88, 439
\bibitem[Landolt (1992)]{lan92} Landolt, A, 1992, \aj, 104, 340
\bibitem[Majewski et al. (1999)]{maj99} Majewski, S.~R., Ostheimer, J.~C., Kunkel, W.~E., Johnston, K.~V., Patterson, R.~J., \& Palma, C.\ 1999, New Views of the Magellanic Clouds, 190, 508
\bibitem[Majewski et al. (2000)]{maj00} Majewski, S. R., Ostheimer, J. C., Kunkel, W. E., \& Patterson, R. J. 2000, \aj, 120, 2550
\bibitem[Majewski et al. (2003)]{Maj03} Majewski, S. R., Strutskie, M. F., Weinberg, M. D., \& Ostheimer, J. C. 2003, \apj, 599, 1082 
\bibitem[Majewski et al. (2009)]{maj09} Majewski, S.~R., Nidever, D.~L., Mu{\~n}oz, R.~R., Patterson, R.~J., Kunkel, W.~E., \& Carlin, J.~L.\ 2009, IAU Symposium, 256, 51
\bibitem[Marigo et al. (2008)]{Mar08} Marigo, P., Girardi, L., Bressan, A., Groenewegen, M.A.T., Silva, L., \& Granato, G.L. 2008, \aap, 482, 883
\bibitem[Massey \& Olsen (2003)]{mas03} Massey, P. \& Olsen, K.~A.~G. 2003, \aj 126, 2867
\bibitem[McClure-Griffiths et al. (2009)]{mcg09} McClure-Griffiths, N. M. et al. 2009, \apjs, 181, 398
\bibitem[Minniti et al. (2003)]{min03}  Minniti, D., Borissova, J., Rejkuba, M., Alves, D. R., Cook, K. H., \& Freeman, K. C. 2003, Science, 301, 1508
\bibitem[Monet et al. (2003)]{mon03} Monet, D. G. et al. 2003, \aj, 125, 984
\bibitem[Morrison et al. (2000)]{mor00} Morrison et al. 2000, \aj, 119, 2254
\bibitem[Morrison et al. (2001)]{mor01} Morrison, H. L., Olszewski, E. W., Mateo, M, Norris, J. E., Harding, P., Dohm-Palmer, R. C., \& Freeman, K. C. 2001, \aj 121, 283
\bibitem[Munoz et al. (2006)]{mun06} Munoz, R. R. et al.  2006, \apj, 649, 201
\bibitem[Nidever et al. (2007)]{nid07} Nidever, D.~L., Majewski, S.~R., Munoz, R.~R., Patterson, R.~J., Kunkel, W.~E., \& Carlin, J.\ 2007, Bulletin of the American Astronomical Society, 38, 942 
\bibitem[Olsen \& Massey (2007)]{ols07} Olsen, K.~A.~G. \& Massey, P. 2007, \apj, 656, 61
\bibitem[Olsen (2009)]{ols09} Olsen, K. 2009, priv. comm.
\bibitem[Olszewski et al. (2009)]{olz09} Olszewski, E., Saha, A., Knezek, P., Subramaniam, A., de Boer, T., \& Seitzer, P. 2009, \aj, 138, 1570
\bibitem[Piatek et al. (2008)]{pia08} Piatek, S., Pryor, C., Olszewski, E. W.  2008, \aj, 135, 1024
\bibitem[Putman et al. (1998)]{put98} Putman, M. E. et al. 1998, Nature, 394, 752
\bibitem[Robin et al. (2003)]{rob03} Robin, A. C., Reyle, C., Derriere, S., \& Picaud. S. 2003, \aap, 409, 523 (erratum: 2004, \aap, 416, 157)
\bibitem[Saha (1985)]{sah85} Saha, A. 1985, \apj, 289, 310
\bibitem[Sandage (1987)]{san87} Sandage, A. 1987, \aj, 93, 610
\bibitem[Sandage \& Fouts (1987)]{sanfou87} Sandage, A. \& Fouts, G. 1987, \aj, 93, 592
\bibitem[Schechter et al. (1993)]{sch93} Schechter, P. L., Mateo, M., \& Saha, A. 1993, \pasp 105, 1342
\bibitem[Schlegel et al.(1998)]{sch98} Schlegel, D.~J.,  Finkbeiner, D.~P., \& Davis, M.\ 1998, \apj, 500, 525
\bibitem[Schommer et al. (1992)]{schom92} Schommer, R. A., Suntzeff, N. B., Olszewski, E. W., \& Harris, H.C. 1992, \aj, 103, 447
\bibitem[Stryker (1984)]{str84} Stryker, L. L. 1984, \apjs, 55, 127
\bibitem[Subramaniam \& Subramanian(2009)]{sub09} Subramaniam, A., \& Subramanian, S.\ 2009, \aap, 503, L9 
\bibitem[Suntzeff et al. (1992)]{sun92} Suntzeff, N. B.,  Schommer, R. A., Olszewski, E. W., \& Walker, A. R. 1992, \aj, 104, 1743
\bibitem[van der Marel (2001)]{vdm01} van der Marel, R. P. 2001, \aj, 122, 1827
\bibitem[van der Marel \& Cioni (2001)]{vdmC01} van der Marel, R. P. \& Cioni, M. L. 2001, \aj, 122, 1807 
\bibitem[van der Marel et al. (2002)]{vdm02} van der Marel, R. P.,  Alves, D. R., Hardy, E., \& Suntzeff, N. B. 2002, \aj, 124, 2639
\bibitem[Weinberg (2000)]{wein00} Weinberg, M. D. 2000, \apj, 532, 922
\bibitem[Weinberg \& Nikolaev(2001)]{wein01} Weinberg, M.~D., \& Nikolaev, S.\ 2001 \apj, 548, 712
\bibitem[Zinn (1985)]{zin85} Zinn, R. 1985, \apj, 293, 424

\end{thebibliography}
\end{document}